\documentclass[aps,prd,twocolumn,showpacs,preprintnumbers,superscriptaddress]{revtex4-1} 
\usepackage{dcolumn} 
\usepackage{graphicx}
\usepackage{slashed,color}
\usepackage{amsmath,amsfonts,bm}
\usepackage{amssymb}
\usepackage{calligra}
\usepackage[T1]{fontenc}

\newcommand{\nn}{\nonumber \\ }

\usepackage{yfonts}
\newcommand{\beq}{\begin{equation}}
\newcommand{\eeq}{\end{equation}}
\newcommand{\bea}{\begin{eqnarray}}
\newcommand{\eea}{\end{eqnarray}}

\def\OMIT#1{{}}

\begin{document}


\title{Lattice QCD exploration of pseudo-PDFs}

\title{Lattice QCD exploration of parton pseudo-distribution functions} 
\author{Kostas Orginos}
\affiliation{Department of Physics, The College of William \& Mary, Williamsburg, VA 23187, USA}
      \affiliation{Thomas Jefferson National Accelerator Facility,
              Newport News, VA 23606, USA}

 \author{Anatoly Radyushkin }
\affiliation{Physics Department, Old Dominion University, Norfolk,
             VA 23529, USA}
       \affiliation{Thomas Jefferson National Accelerator Facility,
              Newport News, VA 23606, USA}
 
  \author{Joseph Karpie } \author{Savvas Zafeiropoulos}
  \affiliation{Department of Physics, The College of William \& Mary, Williamsburg, VA 23187, USA}
      \affiliation{Thomas Jefferson National Accelerator Facility,
              Newport News, VA 23606, USA}

\begin{abstract}

  \noindent
  
  We demonstrate    a new method of extracting
   parton distributions from lattice calculations.  
  The starting idea is to treat the generic equal-time 
   matrix element ${\cal M} (Pz_3, z_3^2)$  as a function 
   of the Ioffe time $\nu = Pz_3$ and the distance $z_3$. The 
   next step  is to  divide ${\cal M} (Pz_3, z_3^2)$  by   the rest-frame density  
    ${\cal M} (0, z_3^2)$.  
 Our lattice calculation  shows  a linear  exponential  $z_3$-dependence 
    in the rest-frame function,  expected from  the $Z(z_3^2)$ factor 
    generated by the gauge link.
 Still, we observe that  the ratio ${\cal M} (Pz_3 , z_3^2)/{\cal M} (0, z_3^2)$
    has a Gaussian-type  behavior with respect to $z_3$ for  6 
    values of $P$ used in the calculation. 
    This means that $Z(z_3^2)$ factor was  canceled  in the ratio. 
When plotted as a function of $\nu$ and $z_3$, the data  
    are very close 
    to $z_3$-independent functions. This  phenomenon corresponds to 
    factorization of the $x$- and $k_\perp$-dependence
    for the  TMD ${\cal F} (x, k_\perp^2)$. 
For small $z_3 \leq 4a$,  the residual  $z_3$-dependence
    is  explained by  perturbative evolution, with  $\alpha_s/\pi =0.1$.

\end{abstract}


\pacs{12.38.-t, 
      11.15.Ha,  
      12.38.Gc  
}

\maketitle


\section{ Introduction}

Extraction of  parton distribution functions (PDFs) $f(x)$ \cite{Feynman:1973xc}
on the lattice is a challenging problem attracting a lot of attention.
 The usual  method to approach PDFs on the lattice  is to calculate  their moments. 
  However, recently, X. Ji  \cite{Ji:2013dva} suggested a method 
 allowing a calculation  of  PDFs   as  functions of $x$. 
 
 Since the PDFs 
are  related to   matrix elements
of bilocal operators on   the light cone $z^2=0$, this was a stumbling block preventing  
a  direct  calculation of these functions in  the   lattice 
 gauge  theory 
 formulated in  Euclidean space. 
 
  To overcome this difficulty, X. Ji proposes to use purely space-like
separations $z=(0,0,0,z_3)$.  
The functions in this case are   quasi-PDFs  $Q(y,p_3)$ describing the distribution 
of   the $p_3$  hadron  momentum  component.
The key point is that quasi-PDFs $Q(y,p_3)$
   tend to  usual PDFs $f(y)$ 
     in the \mbox{$p_3 \to \infty$}  limit.
     The same method can be applied to distribution amplitudes (DAs).  
The results of  quasi-PDF calculations on the  lattice  
   were reported  
in Refs.  \cite{Lin:2014zya,Chen:2016utp,Alexandrou:2015rja}
and of the pion quasi-DA   in Ref. \cite{Zhang:2017bzy}.

Recent papers \cite{Radyushkin:2016hsy,Radyushkin:2017gjd}
by one of the authors (A.R.) contain an investigation of the  nonperturbative $p_3$-evolution of
quasi-PDFs and quasi-DAs. This study  is based on 
 the formalism of  virtuality distribution functions 
 \cite{Radyushkin:2014vla,Radyushkin:2015gpa}.
 The approach developed in Refs.
 \cite{Radyushkin:2016hsy,Radyushkin:2017gjd}  has  established  a connection between 
the   quasi-PDFs  and 
  the   ``straight-link'' transverse momentum dependent 
 distributions    (TMDs) ${\cal F} (x, k_\perp^2)$. 
 Starting  from  simple models for TMDs,  models were 
 built  for the    nonperturbative evolution of
quasi-PDFs.  The derived curves agree qualitatively with the patterns of
\mbox{$p_3$-evolution}  produced by lattice simulations.    

The structure  of quasi-PDFs was further studied in Ref.   \cite{Radyushkin:2017cyf}.
 It was   shown  that, 
when a  hadron is moving, the parton $k_3$ momentum
may be treated  as 
coming  from two sources. The hadron's motion as a whole yields 
the  $xp_3$  part,  which is governed by the dependence of
 the TMD  ${\cal F} (x, \kappa^2)$ on its first argument namely $x$.
  The residual part $k_3-xp_3$   
is controlled by the  way that the TMD depends on its second argument,
$\kappa^2$, 
which dictates the shape of the primordial 
 rest-frame 
 momentum distribution.    Quasi-PDFs due to their convolution nature possess a rather  involved pattern of their 
 \mbox{$p_3$-evolution,} making mandatory relatively big values $p_3 \gtrsim 3$ GeV in order to safely approach the PDF limit.
 
To accelerate the convergence,   a different  approach for the 
PDF extraction from lattice  calculations 
 was  proposed  \cite{Radyushkin:2017cyf}. It is  based on the concept of 
{\it pseudo-PDFs} $ {\cal P} (x, z_3^2) $.  They  generalize 
the light-cone PDFs $f(x)$ onto spacelike intervals like  
 $z=(0,0,0,z_3)$.   The pseudo-PDFs are Fourier transforms 
 of the  \mbox{{\it Ioffe-time}  \cite{Ioffe:1969kf}}  \mbox{{\it distributions} \cite{Braun:1994jq}} 
 ${\cal M} (\nu, z_3^2)$ 
 which are generically given by matrix elements  $  \langle p |   \phi(0) \phi (z)|p \rangle $
 written as  functions of $\nu = p_3 z_3$ and $z_3^2$. 
In contrast to quasi-PDFs, the pseudo-PDFs have  the ``canonical''  $-1 \leq x  \leq 1$ support
 for all values of $z_3^2$.  In the limit  $z_3\to 0$ they tend to PDFs, 
  showing, in this  limit, a typical perturbative evolution 
 with the scale $1/z_3$ being the parameter of evolution.

{As discussed in~ \cite{Radyushkin:2016hsy,Radyushkin:2017gjd},  the  fast  nonperturbative
decrease  with $z_3^2$ of   the pseudo-PDFs  $ {\cal P} (x, z_3^2) $  or the 
Ioffe-time distribution  ${\cal M} (\nu, z_3^2)$,  is responsible for   delaying the approach of quasi-PDFs  $Q(y, p_3)$ 
to the  PDF  $f(y)$}.
 An important observation is that one can  strongly 
 reduce the   $z_3^2$-dependence
by simply dividing  the Ioffe-time distribution 
${\cal M} (\nu, z_3^2)$ by an appropriate  factor $D(z_3^2)$ satisfying $D(0)=1$
and having the $z_3^2$-dependence close (on average) to that of
${\cal M} (\nu, z_3^2)$. 
The absence of  the \mbox{$\nu$-dependence}  in this factor and its $D(0)=1$ 
normalization guarantees that   the ratio ${\cal M} (\nu, z_3^2)/D(z_3^2)$ 
taken in the $z_3^2 \to 0$ limit  will produce the same PDF as the original function
${\cal M} (\nu, z_3^2)$ taken in the same limit.

The choice for  $D(z_3^2)$ advocated in Ref.   \cite{Radyushkin:2017cyf},
is   to take it to  be equal to  the rest-frame function $ {\cal M} (0, z_3^2)$. 
An  additional advantage of this choice  is that 
both ${\cal M} (\nu, z_3^2)$ and ${\cal M} (0, z_3^2)$ contain the same multiplicative 
factor $Z(z_3^2)$ generated by the  renormalization of the gauge link.
In the ratio, it should cancel out. 

Our  goal in the present work  is an exploratory   lattice calculation 
of the $u$-$d$ proton PDF using the strategy outlined in \mbox{Ref.   \cite{Radyushkin:2017cyf}. } 
 To make this article  self-contained, we reproduce in Sections II and III  the main ideas 
 of \mbox{Ref.   \cite{Radyushkin:2017cyf}.}   The description of the method   used for the  lattice extraction of the 
 reduced Ioffe-time distribution is given in Section IV. 
 The data analysis   and interpretation is discussed  in Section V.
 The summary of the paper is given in Section VI.

 \section{Parton distributions} 


  \subsection{Generic matrix element  and 
parton distributions
} 

 The basic object for defining parton distributions is a matrix element 
 of a bilocal operator that  (skipping inessential details of  its 
 spin structure)   may be written   generically  like 
 $\langle  p | \phi (0)  \phi(z)  | p \rangle $. 
Due to invariance under Lorentz transformations, it is given by a function of two scalars,  $(pz)$
(which will  be denoted by $-\nu$)  and   $z^2$
(or $-z^2$, in order to have a positive value  for \mbox{spacelike $z$)}
  \begin{align}
  \langle p |   \phi(0) \phi (z)|p \rangle 
=  & {\cal M} (-(pz), -z^2)  =  {\cal M} (\nu, -z^2)  
\,  .
 \
 \label{lorentz}
\end{align} 
 One can demonstrate \cite{Radyushkin:2016hsy,Radyushkin:1983wh}    that, for all relevant Feynman diagrams,   
 its  Fourier transform  ${\cal P} (x, -z^2)$ with respect to $(pz)$ 
 has $-1 \leq x \leq 1$ as support, i.e., 
   \begin{align}
 {\cal M} (-(pz), -z^2) 
&   = 
 \int_{-1}^1 dx 
 \, e^{-i x (pz) } \,  {\cal P} (x, -z^2)  \   .
  \label{MPD}
\end{align}   
  Eq. (\ref{MPD}) serves as a covariant definition of $x$.  
   In this  definition of $x$, one does not need to assume that 
$p^2=0$ or $z^2=0$.


Choosing a light-like  $z$, e.g., 
having  solely the light-front component $z_-$,  we   
parametrize the matrix element  by $f(x)$, the twist-2 parton distribution  
  \begin{align}
  {\cal M} (-p_+ z_- , 0)  =
   \int_{-1}^1 dx \, f(x) \, 
e^{-ixp_+ z_-} \,  \   . 
 \label{twist2par0}
\end{align}
One can  rewrite this definition as 
  \begin{align}
  {\cal M} (\nu, 0)  =
   \int_{-1}^1 dx \, f(x) \, 
e^{ix\nu} \,  \  . 
 \label{twist2par1}
\end{align}
The inverse relation is given by 
\begin{align} 
 f  (x)  =\frac{1}{2 \pi}  \int_{-\infty}^{\infty}  d\nu \, e^{-i x \nu}  \, {\cal M}(\nu , 0)    = {\cal P} (x, 0) 
  \  .
\label{fxMnu}
\end{align}  

Due to the fact that $f(x) =   {\cal P} (x, 0) $, the function $ {\cal P} (x, -z^2) $ provides a generalization of the concept of 
PDFs onto non-lightlike intervals $z^2$ (in principle, $z^2$ may be even timelike).
 Following   \cite{Radyushkin:2017cyf} ,  we will be referring to it as the
 {\it pseudo-PDF}.  The variable $(pz)=-\nu $ is  called  often  the {\it Ioffe time} \cite{Ioffe:1969kf}, 
and  consequently ${\cal M}(\nu , -z^2) $  is  the {\it Ioffe-time distribution} \cite{Braun:1994jq}.

  In renormalizable theories (including QCD),    
  the function  $ {\cal M} (\nu, -z^2)  $ 
 has  logarithmic $\sim \ln (-z^2)$  singularities 
which generate the perturbative evolution of parton densities.  
In the approach based on  the operator product expansion
(OPE),  the standard procedure is to  remove  these singularities  with the help of some 
 prescription. 
The most popular of them is the  $\overline{\rm MS}$  scheme based 
on dimensional  regularization.  
Consequently the resulting PDFs  have a dependence on the renormalization scale 
$\mu$, and therefore one should write the PDFs as $   f(x, \mu^2)$.  

At small spacelike $z^2$ and at the leading logarithm level, the pseudo-PDFs are 
related to the $\overline{\rm MS}$ distributions by a simple rescaling of their second arguments.  
In particular, when $z^2=-z_3^2$, one has 
\begin{align} 
    {\cal P} (x, z_3^2) = f \left (x, (2 e^{-\gamma_E}/ z_3)^2 \right )
  \  ,
\label{Ptof}
\end{align}  
where $\gamma_E$  is the Euler's constant.  
The rescaling  factor  between $\mu$ and $1/z_3$ is very close to 1, since \mbox{$2e^{-\gamma_E}=1.12$. }

 \subsection{Transverse momentum dependent-  and quasidistributions}

Treating the target momentum $p$ as   longitudinal,
\mbox{$p= (E, {\bf 0}_\perp, P)$,}  one can introduce  
transverse degrees of freedom. 
In particular, taking $z$ that has 
 $z_-$ and \mbox{$z_\perp= \{z_1,z_2\}$}  components only,  one defines  the  
{\it TMD }   ${\cal F} (x, k_\perp^2)$   
       \begin{align}
 {\cal P} (x,  z_\perp^2) 
&   = 
    \int  d^2{\bf k}_\perp   e^{i ({\bf k}_\perp {\bf z}_\perp)} 
      {\cal F} (x, {k}_\perp^2)
 \   .
  \label{MTMD0}
\end{align}  
In  this context, the pseudo-PDFs  $ {\cal P} (x,  z_\perp^2) $ actually coincide
with the  {\it impact parameter distributions},   a familiar object 
used in many TMD studies.




Since one cannot arrange  light-like separations on the lattice,
it was proposed  \cite{Ji:2013dva}  to consider equal-time 
spacelike separations  
$z= (0,0,0,z_3)$ (or, for brevity, \mbox{$z=z_3$}). Then, 
 in  the   \mbox{$p=(E, 0_\perp, P)$}  frame,
 one  can introduce the quasi-PDF  $Q(y, P)$    through a parametrization  
  \begin{align}
  \langle p |   \phi(0) \phi (z_3)|p \rangle 
=  & 
\int_{-\infty}^{\infty}   dy \, 
 Q(y, P) \,  e^{i y  P z_3 } \, 
 \  . 
 \label{newVDFxzQ}
\end{align} 
According to  this definition, the  quasi-PDF  $Q(y,P)$ describes  the probability 
that  the  parton carries the  fraction $y$ of the parent hadron's third momentum
component $P$.  
The variables $\nu$ and $-z^2$
 in this case are given by $Pz_3$ and $z_3^2$, so we have
 \begin{align}
{\cal M} (\nu,z_3^2) 
=  & 
\int_{-\infty}^{\infty}   dy \, 
 Q(y, P) \,  e^{i y  \nu } \, 
 \  . 
 \label{newVDFxzQ}
\end{align} 
Since $z_3^2= \nu^2/P^2$,   the inverse  Fourier transformation 
may be written as 
\begin{align} 
  Q(y,  P)   =\frac{1}{2 \pi}  \int_{-\infty}^{\infty}  d\nu \, 
   \, e^{-i y  \nu}  \, {\cal M} (\nu,  \nu^2/P^2)    \  .
\label{IxM}
\end{align}  
It 
shows that $  Q(y,  P)  $  tends to $f(y)$  in the 
$P \to \infty$  limit,  since  formally    ${\cal M} (\nu,  \nu^2/P^2) \to {\cal M} (\nu, 0)$
when \mbox{$P \to \infty$. }


As established  in Ref.  \cite{Radyushkin:2016hsy}, quasi-PDFs may be written in terms of TMDs %
  \begin{align}
 Q(y, P) /P =  & \,\int_{-\infty}^{\infty} d  k_1
\int_{-1} ^ {1} d  x \,   {\cal F} (x, k_1^2+(y-x)^2P^2 )
  \  . 
 \label{QTMD}
 \end{align} 
  
\subsection{Quantum chromodynamics (QCD)  case}  

In case of %
  the non-singlet parton densities of QCD,
one is considering matrix elements  
    \begin{align}
 {\cal M}^\alpha  (z,p) \equiv \langle  p |  \bar \psi (0) \,
 \gamma^\alpha \,  { \hat E} (0,z; A) \psi (z) | p \rangle \  , 
\label{Malpha}
\end{align}
 where  $
{ \hat E}(0,z; A)$ is  the  standard  $0\to z$ straight-line gauge link 
 in the quark (fundamental) representation.
 These matrix elements can  be decomposed into $p^\alpha$ and $z^\alpha$ parts
\begin{align} 
{\cal M}^\alpha  (z,p) = &2 p^\alpha  {\cal M}_p (-(zp), -z^2) 
 + z^\alpha  {\cal M}_z (-(zp),-z^2)
\ .
\end{align}
Only %
the ${\cal M}_p (-(zp), -z^2) $ part gives the twist-2 distribution when  $z^2 \to 0$.

Introducing TMDs,   one  takes   $z=(z_-, z_\perp)$  and  the \mbox{$\alpha=+$}  component of 
${\cal M}^\alpha$. Hence,  the  $z^\alpha$-part drops out.
After that,  $ {\cal M}_p(\nu, z_\perp^2)$ is   the only 
surviving part of ${\cal M}^\alpha  (z,p)$, and in the remaining discussion 
we use the short hand notation of ${\cal M} \equiv {\cal M}_p$.

In the case of    quasidistributions $Q(y,P)$,  
we   can  avoid  the $z^\alpha$ contamination by   considering   the
  time component of \mbox{$ {\cal M}^\alpha  (z=z_3,p)$}  and defining 
 \begin{align}
&  {\cal M}^0   (z_3,p)   
  =  2p^0 
 \int_{-1}^1 dy\,  
Q(y,P) \, 
 \,  e^{i  y Pz_3 }  \  . 
 \label{OPhixspin12}
\end{align} 



\subsection{Factorized models} 
 
 The structure of the quasi-PDFs may be illustrated on the example
 of the simplest models in which 
the nonperturbative (or soft) part of the TMDs 
 ${\cal F}  (x, k_\perp^2)$ is  represented   by a product 
    \begin{align}
{\cal F}^{\rm soft}  (x, k_\perp^2) = f(x) K(k_\perp^2)
\end{align} 
 of the 
 collinear parton distribution $f(x)$  and a \mbox{$k_\perp^2$-dependent}  factor $K(k_\perp^2)$, usually 
 modeled by a Gaussian.
 As we shall see, the quasi-PDFs  have a rather complicated structure,
 even when they  are built from   these simple  factorized models.
 
\begin{figure}[t]
    \centerline{\includegraphics[width=3.1in]{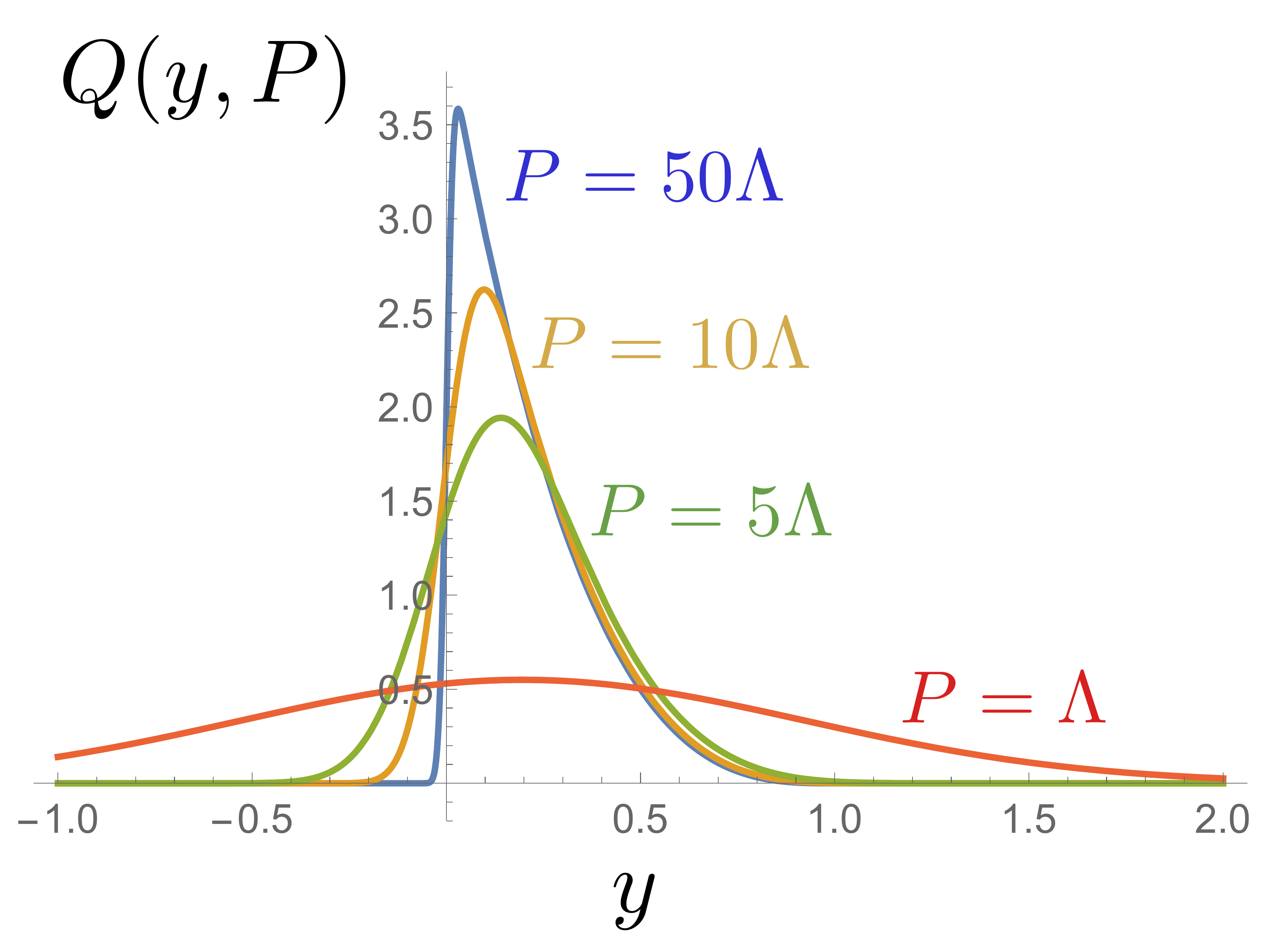}}
    \caption{Evolution of quasi-PDF $Q(y,P)$  in the  factorized Gaussian  model  for {$P/\Lambda =1, 5, 10, 50$}.
    \label{Qgy}}
    \end{figure}

For  the Ioffe-time distribution  ${\cal M} (\nu,-z^2)$,  this Ansatz corresponds 
to the factorization assumption 
   \begin{align}
   {\cal M}^{\rm soft}  (\nu,z_3^2) = {\cal M}^{\rm soft}  (\nu,0){\cal M} (0,z_3^2) \ 
    \end{align} 
    for its soft part. 
Still, even if the soft TMD factorizes,  the soft part of the  quasi-PDF has the  convolution
structure of Eq. (\ref{QTMD}).    Taking,   for example,   a Gaussian  form  
  \begin{align}
K_G(k_\perp^2) =  & 
\frac{1}{\pi  \Lambda^2}  e^{-k_\perp^2/\Lambda^2}
  \  , 
 \label{DTMDsmallfac}
\end{align}  
one gets the following model  for the quasi-PDF
 \begin{align}
 Q_G(y, P)  = &\frac{P}{\Lambda \sqrt{\pi} }  \,
 \int_{-1}^1 dx\,  %
f(x) \, 
  e^{- (x -y)^2 P^2 / \Lambda^2 }
 \  . 
 \label{QinG}
\end{align} 
Choosing for $f(x)$ 
 a simple toy PDF resembling  the nucleon valence densities
$f(x)=4(1-x)^3 \theta (0\leq x \leq 1)$, one gets the curves shown in Fig. 
 \ref{Qgy}. 
 For large $P$, the quasi-PDF clearly tends to the $f(y)$   PDF  form.
However, only for   
$P \sim 10 \Lambda$  one gets  a quasi-PDF that is rather close to the $P \to \infty$ limiting shape. 
Still, since  $\Lambda \sim \langle k_\perp \rangle$,    
one   translates  the 
 \mbox{$P\sim 10 \Lambda$}  estimate into $P \sim 3$ GeV,
which is  rather  large. 

\section{Pseudo-PDFs}  

The involved  structure  of a quasi-PDF
$Q(y,P)$  can be attributed to the formal  fact that it is given by the Fourier $\nu$-transform  of 
the function 
${\cal M} (\nu, \nu^2/P^2)$,  in which    $\nu$ appears  both in the first and second 
argument of the Ioffe-time distribution.  
One should take   $P$-values that   are sufficiently large 
to neglect the $\nu$-dependence coming from the second argument.

Another way  \cite{Radyushkin:2017cyf} is to try to {\it eliminate}  the \mbox{$z_3^2$-dependence} 
induced by ${\cal M} (\nu, z_3^2)$. The main  idea is based
on the observation  that if one takes the 
$\nu$-Fourier transform  of the modified function 
${\cal M} (\nu, z_3^2)/D(z_3^2)$,  the $z_3\to 0$  limit 
will   give the same PDF as the original Ioffe-time distribution,
provided that   $D(z_3^2)$ is a function of $z_3^2$ only
(but not of $\nu$) and is  equal to 1 for \mbox{$z_3^2=0$.} 

Thus,  the strategy is to  find a function $D(z_3^2)$ whose \mbox{$z_3^2$-dependence } 
would compensate, as much as possible,  the  $z_3^2$-dependence 
of  ${\cal M} (\nu, z_3^2)$.
The next step is to fit the residual polynomial $z_3^2$-dependence 
by polynomials of $z_3^2$ (they may be different for different values of $\nu$),
and in this way extrapolate the data to $z_3^2=0$ limit.
The Fourier transform
of  the resulting function  would correspond to 
 the same PDF  as the $z_3^2$ limit of the original Ioffe-time  distribution  
 ${\cal M} (\nu, z_3^2)$.

In the most lucky situation,  the ratio ${\cal M} (\nu, z_3^2)/D(z_3^2)$ 
would have no polynomial $z_3^2$-dependence  (or just a very mild one). 
In particular, when 
${\cal M} (\nu, z_3^2)$ factorizes, i.e.,  
 ${\cal M} (\nu, z_3^2)= {\cal M} (\nu, 0) {\cal M} (0, z_3^2)$, one should take     
  \mbox{$D(z_3^2) ={\cal M} (0, z_3^2)$}.  In this case,   
 the reduced function
 \begin{align}
{\mathfrak M} (\nu, z_3^2) \equiv \frac{ {\cal M} (\nu, z_3^2)}{{\cal M} (0, z_3^2)} \  
 \label{redm}
\end{align}
is equal to $ {\cal M} (\nu, 0)$, 
and   
the  task  of obtaining  the $z_3\to 0$ limit is accomplished.   

While there is no ``first principle'' reason for such a factorization,
one may expect that  the functions ${\cal M} (\nu, z_3^2)$  
for different $\nu$ have more or less similar dependence on $z_3$,
basically 
reflecting the finite size of the nucleon.

As we mentioned already, the soft part of ${\cal M} (\nu, z_3^2)$  factorizes
if  the soft part of TMD ${\cal F} (x,k_\perp^2)$ factorizes.  
That this happens, is
a standard  assumption of the TMD practitioners
(see, e.g., Ref.  \cite{Anselmino:2013lza}). 
So, there are good chances that  this part of the 
$z_3^2$-dependence of ${\cal M} (\nu, z_3^2)$ 
will be canceled or strongly reduced by the rest-frame function ${\cal M} (0, z_3^2) $.

On the lattice, there is another (and troublesome, see, e.g., Ref. \cite{Ishikawa:2016znu}) source 
of $z_3$-dependence: 
 the  $Z(z_3^2)$ factor
 generated 
by the   renormalization of 
the gauge link  ${ \hat E} (0,z_3; A)$. 
Fortunately, this  problematic     factor $Z(z_3^2)$ 
does not depend on $\nu$ and 
is   the same for  the numerator and denominator 
of the ratio  ${\mathfrak M} (\nu, z_3^2) $. 
This provides another motivation for using ${\cal M} (0, z_3^2)$ 
as a factor $D(z_3^2)$. 

 Thus, the proposal is to perform a lattice study 
 of the reduced Ioffe-time  function ${\mathfrak M} (\nu, z_3^2)$.
 Even if it would have a residual polynomial $z_3^2$-dependence,
 it should  be much  easier to  extrapolate  this dependence to $z_3=0$,
 than the $z_3$-dependence of the original Ioffe-time  distribution
 $ {\cal M} (\nu, z_3^2)$.

Furthermore,   if one observes that  the ratio 
$ {\mathfrak M} (\nu, z_3^2)$ does not have $z_3$-dependence, 
one should  conclude that $ {\cal  M} (\nu, z_3^2)$ factorizes.   
In fact, such a factorization has been already observed 
several years ago 
in the pioneering study \cite{Musch:2010ka} of the 
transverse momentum distributions in  lattice QCD.

Still,  there is an unavoidable source of factorization breaking.
When $z_3$  is small,
$ {\cal M} (\nu, z_3^2)$ has   logarithmic $\ln z_3^2$ singularities
generating the perturbative  evolution of PDFs. 
As we discussed,  
 $1/z_3$ 
 is analogous then  to the renormalization parameter $\mu$ 
of  the scale-dependent PDFs $f(x,\mu^2)$ within the standard OPE approach.  

    \begin{figure}[t]
   \centerline{ \includegraphics[width=3.3in]{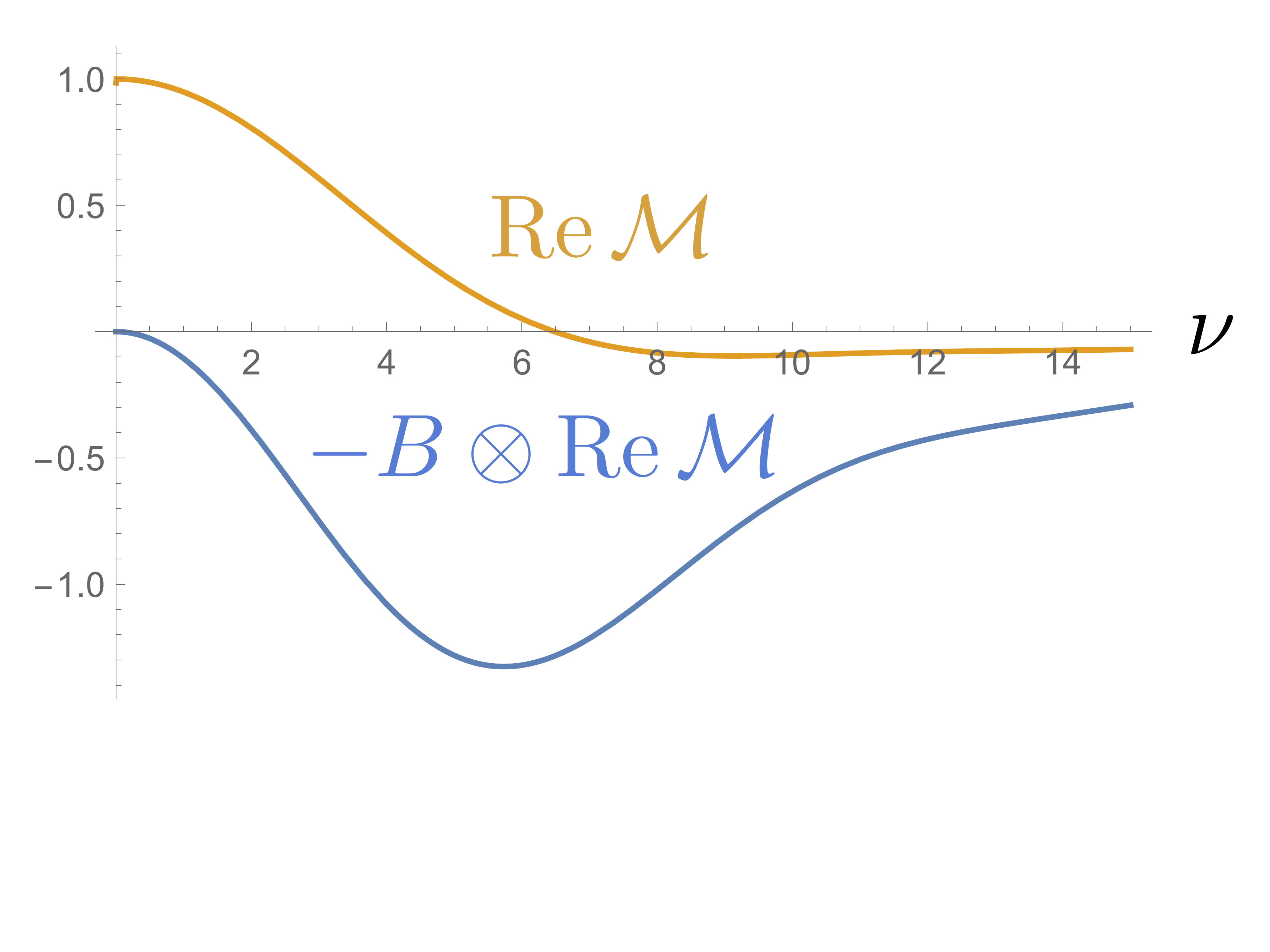}  }
    \vspace{-6mm}
    \caption{Real  part   of model  distribution ${\cal M} (\nu)$  and
    the function $-B \otimes  {\rm Re} \,  {\cal M}$   that  governs its  evolution
    (the minus sign here is   for convenience of  placing  two curves on one   figure).
    \label{MBM}}
    \end{figure}

More specifically, for small values of $z_3$,  the  pseudo-PDF ${\cal P} (x,z_3^2)$
satisfies a  leading-order evolution equation   
 with respect to $1/z_3$  that 
is identical to  the evolution equation for $f (x,\mu^2)$ with respect to $\mu$. 
The  evolution equation  
for  the
reduced %
 Ioffe-time distribution ${\mathfrak M} (\nu, z_3^2)$ can also be written    \cite{Radyushkin:2017cyf} 
    \begin{align}
    \frac{d}{d \ln z_3^2} \,  
{\mathfrak M} (\nu, z_3^2)    &= - \frac{\alpha_s}{2\pi} \, C_F
\int_0^1  du \,   B ( u )   {\mathfrak M} (u \nu, z_3^2)  
 , 
\label{EE}
 \end{align}
where  $C_F=4/3$, and the leading-order evolution kernel $B(u)$ for the non-singlet quark case
is given  \cite{Braun:1994jq} by
   \begin{align}
B (u)    &=  \left [ \frac{1+u^2}{1- u} \right ]_+  \ 
 , 
\label{Bu}
 \end{align} 
 where $[ \ldots ]_+$  denotes the  ``plus'' prescription, i.e. 
    \begin{align}
& \int_0^1  du \,    \left [ \frac{1+u^2}{1- u} \right ]_+     {\mathfrak M} (u \nu)  
\nn & =
\int_0^1  du \,  \frac{1+u^2}{1- u} \,    [{\mathfrak M} ( \nu)-{\mathfrak  M} (u \nu)  ]\
 . 
\label{plus}
 \end{align}

 Note that being a Fourier  transform,
  \begin{align}
  {\cal M} (\nu)  =
   \int_{-1}^1 dx \, f(x) \, 
e^{ix\nu} \,  \  , 
 \label{Mf}
\end{align}
 the Ioffe-time distribution has real and imaginary parts even if the function $f(x)$
 is real (which  is the case with parton distributions). In particular, 
   \begin{align}
  {\rm Re} \, {\cal M} (\nu)  =
   \int_{-1}^1 dx \, f(x) \, 
\cos (x\nu)  \,  \  , 
 \label{ReMf}
\end{align}
 and 
    \begin{align}
  {\rm Im} \, {\cal M} (\nu)  =
   \int_{-1}^1 dx \, f(x) \, 
\sin (x\nu)  \,  \  . 
 \label{ImMf}
\end{align}

\begin{figure}[t]
     \centerline{ \includegraphics[width=3in]{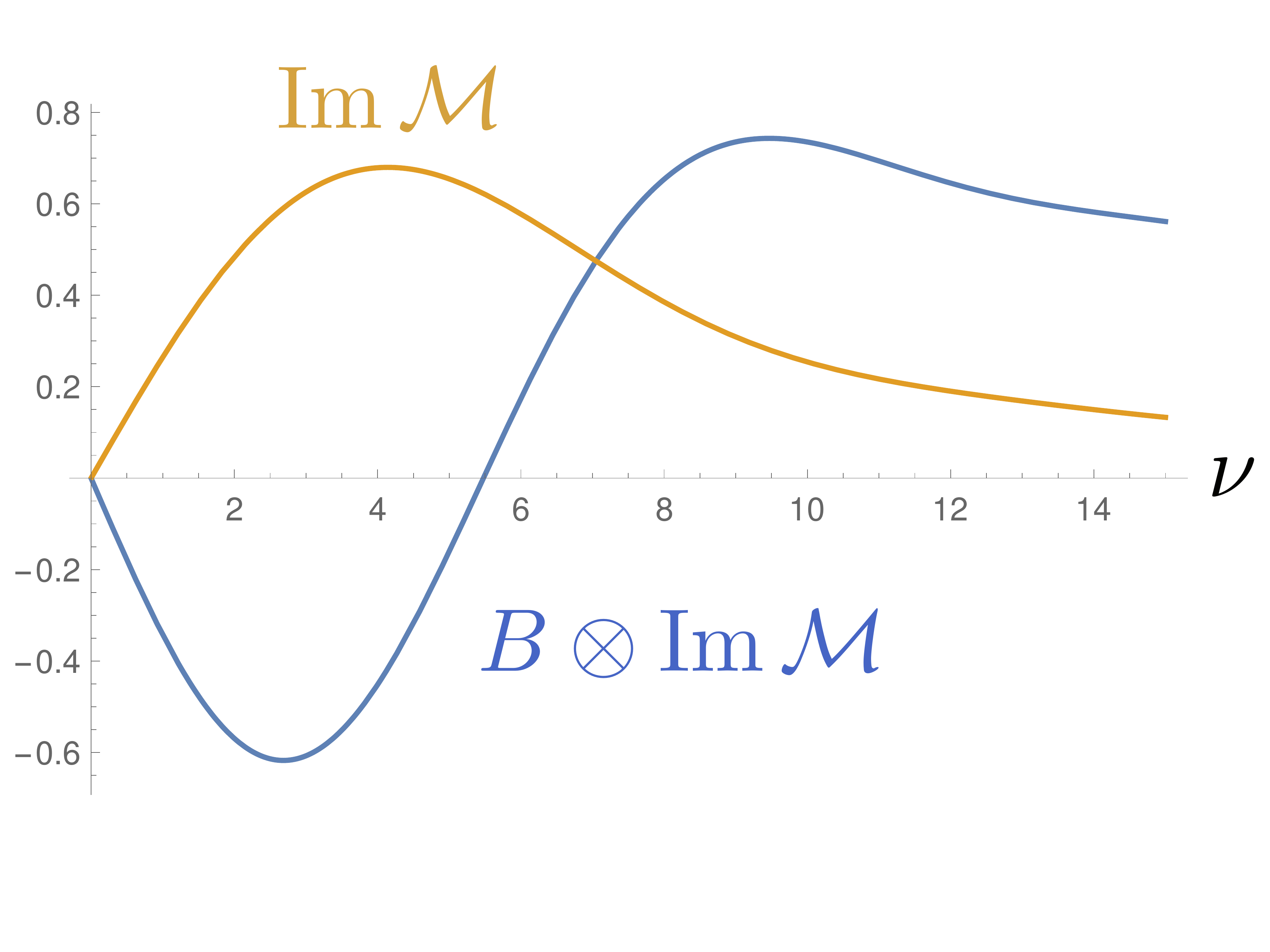}}
    \caption{Imaginary  part of model Ioffe-time distribution $  {\cal M} (\nu)$  and
    the function   $B \otimes  {\rm Im} \,  {\cal M}$  that governs its   evolution.
    \label{MBMI}}
    \end{figure}

  In \mbox{Fig. \ref{MBM}},  we show the function 
  ${\rm Re}  \, {\cal M} (\nu)$ for a model PDF 
      \begin{align}
q (x)= \frac{315}{32}  \sqrt{x} (1-x)^3 \theta ( 0\leq x \leq 1) \ .
\end{align}
Its integral is normalized to 1, and it  is nonzero for positive $x$ only, 
  which corresponds  to  the absence of antiquarks. 
As we shall see, this particular form    appears in the description of   actual  lattice data.
In \mbox{Fig. \ref{MBMI}}, we     show the  function 
  ${\rm Im}  \, {\cal M} (\nu)$ for the same  model PDF.

  We also show in these figures 
the convolution integrals governing the  evolution, namely 
\mbox{$-B \otimes {\rm Re}\, {\cal M}(\nu)$}     
and \mbox{$B \otimes {\rm Im}\, {\cal M}(\nu)$}.     
The reader can notice that  $B \otimes {\rm }\,  {\cal M}(\nu)$  is zero for 
$\nu=0$,  resulting from the vector current conservation.
As a consequence, the perturbative evolution leaves the rest-frame density ${\cal M} (0,z_3^2)$  (which is always real) unaffected.  In other words, the $\ln z_3^2$ terms 
are present  only  in the numerator ${\cal M} ( \nu, z_3^2)  $ 
of the $ {\mathfrak M} (\nu, z_3^2)$ ratio,
but not in its ${\cal M} ( 0, z_3^2)  $  denominator.  

Note also that  the evolution of the real part always leads 
to a {\it decrease} of   ${{\rm Re} \, \cal M} ( \nu, z_3^2)$ when $z_3^2$ increases.
For the imaginary part, the evolution pattern is more complicated.
Namely,    below $\nu \sim 5.5$, the function  ${{\rm Im} \, \cal M} ( \nu, z_3^2)$
{\it increases} 
 when $z_3^2$ increases. Only above $\nu \sim 5.5$, the evolution leads 
 to a decrease of ${{\rm Im} \, \cal M} (u \nu, z_3^2)$ with $z_3^2$,
 and  the evolution pattern becomes  similar to that of the real part.

\section{Numerical investigation}
\label{sec:numerical}

  In order to check numerically the  ideas discussed above we performed lattice 
  QCD calculations in the quenched approximation at $\beta=6.0$ on $32^3\times 64 $
  lattices (lattice spacing $a=0.093$ fm). We used the non-perturbatively tuned clover fermion action with the
   clover coefficients computed by the Alpha  \mbox{collaboration \cite{Luscher:1996ug}.}
  
   We used a total of 500 configurations separated by 1000 updates each one consisting of four over-relaxation   and one 
   heatbath sweeps. 
  On each configuration we computed correlation functions from 6 randomly selected point sources.
   The pion  and nucleon masses in this setup were determined to be $601(1)$ MeV  
   and $1411(4)$MeV respectively. Conversion to physical energy units was 
   performed used the Alpha collaboration scale setting 
   for quenched QCD \cite{Necco:2001xg}. 
   
\begin{figure}[t]
  \includegraphics[width=0.45\textwidth]{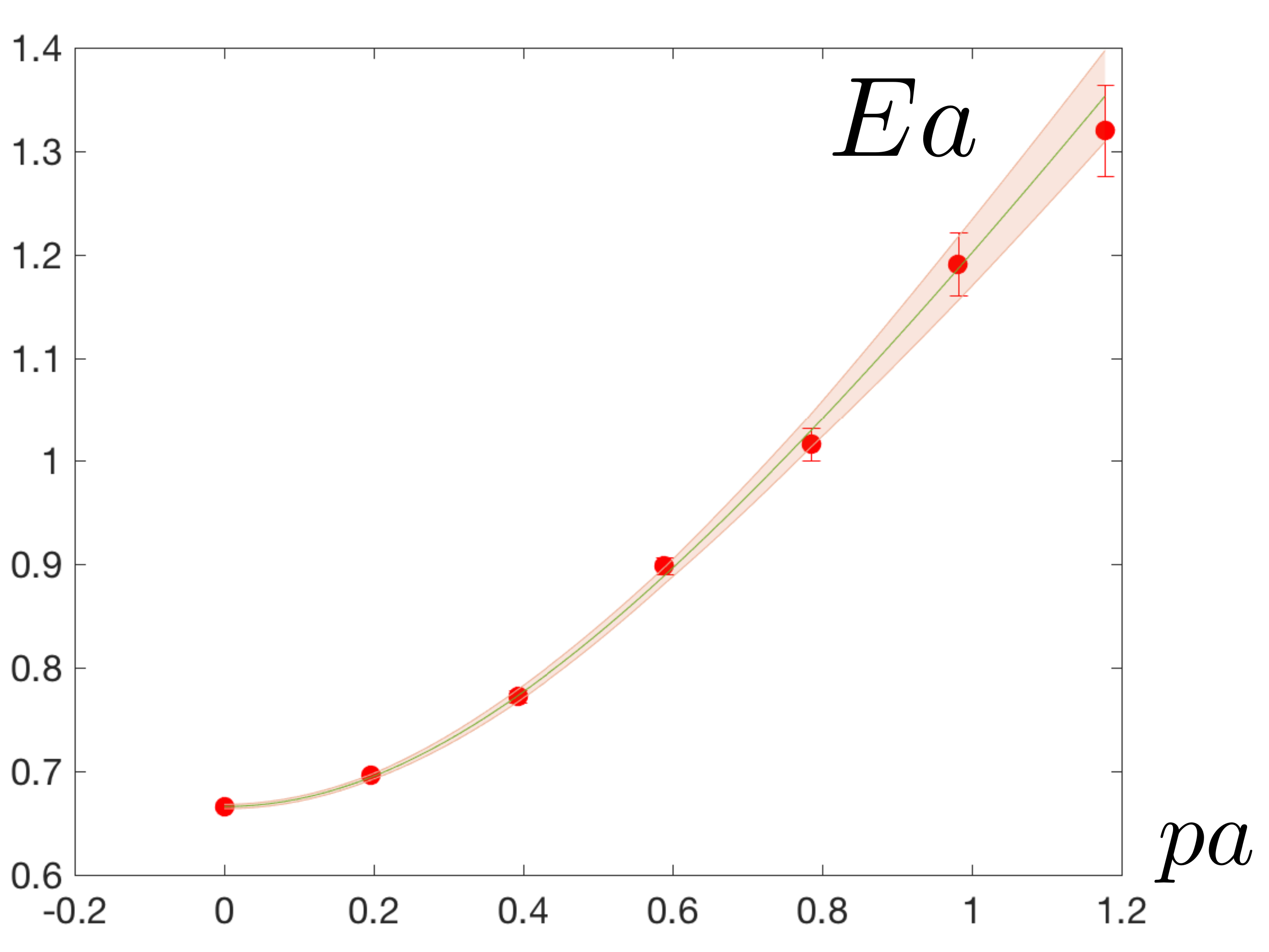} 
  \caption{Nucleon dispersion relation. Energies and momenta are in lattice units. The solid line is the continuum dispersion relation (not a fit) while the errorband is an indication of the statistical error of the lattice nucleon energies. }
  \label{fig:disp}
\end{figure}

   Our nucleon states were boosted up to a total momentum of $2.5\, $GeV (corresponding to the 6th lattice momentum).
   Inside  this momentum  range, the continuum dispersion relation for the nucleon was satisfied within the errors of the calculation,
    indicating small lattice artifacts of ${\cal O}(aP)$.
    In Fig. \ref{fig:disp} we plot the nucleon energy as a function of   momentum along with the continuum dispersion relation corresponding to our lattice nucleon zero momentum energy. 
    
  The computation of the matrix elements was performed using 
   the methodology described in~\cite{Bouchard:2016heu} with an operator insertion given by Eq.~(\ref{Malpha}). 
  Taking the time component of the current we can isolate  $\mathcal{M}_p(-z\cdot p, -z^2)$ which as discussed above is directly related to PDFs.
  
  Following~\cite{Bouchard:2016heu} we need to compute two types of correlation functions. The first is a regular nucleon two point function given by
  \begin{equation}
  C_P(t) = \langle \mathcal{N}_P(t)\overline{ \mathcal{N}}_P(0) \rangle\,  ,
  \end{equation}
where $ \mathcal{N}_P(t)$ is a helicity averaged, non-relativistic nucleon interpolating field with momentum $p$. The quark fields in $ \mathcal{N}_p(t)$ are smeared with a gauge  invariant Gaussian smearing. This choice of an interpolation field 
is known to couple well to the nucleon ground state (see discussion in~\cite{Bouchard:2016heu}).
 The quark smearing width was optimized to give good  {overlap with the nucleon ground state within the}  range of momenta {in our calculation}.  The second correlator is given by
  \begin{equation}
  C^{\mathcal{O}^0(z)}_P(t) =\sum_\tau \langle \mathcal{N}_P(t) \mathcal{O}^0(z,\tau)\overline{ \mathcal{N}}_P(0) \rangle\, ,
  \end{equation}
  where 
   \begin{equation}
  \mathcal{O}^0(z,t)= \overline\psi(0,t) \gamma^0 \tau_3 \hat{E}(0,z;A)\psi(z,t)\,,
  \end{equation}
 with $\tau_3$  being the flavor Pauli matrix.
  The proton momentum and the displacement of the quark fields were both taken along the $\hat z$ axis ($\vec z = z_3 \hat z$ and $\vec p = P \hat z$).
  We  define the effective matrix element as 
  \begin{equation}
  \mathcal{M}_{\rm eff}(z_3 P, z_3^2;t) = \frac{C^{\mathcal{O}^0(z)}_P(t+1)}{C_P(t+1)} - \frac{C^{\mathcal{O}^0(z)}_P(t)}{C_P(t) }  \  . 
  \end{equation}
  As it was shown in~\cite{Bouchard:2016heu}, our matrix element $\mathcal{J}$ can then be extracted at the large Euclidean time separation as
  \begin{equation}
\frac{\mathcal{J}(z_3P, z_3^2)}{2 E}= \lim_{t\rightarrow\infty}  \mathcal{M}_{\rm eff}(z_3 P, z_3^2;t) \  ,
  \end{equation} 
  where $E$ is the energy of the nucleon. 
       \begin{figure}[t]
       \vspace{-6mm}
  \includegraphics[width=0.45\textwidth]{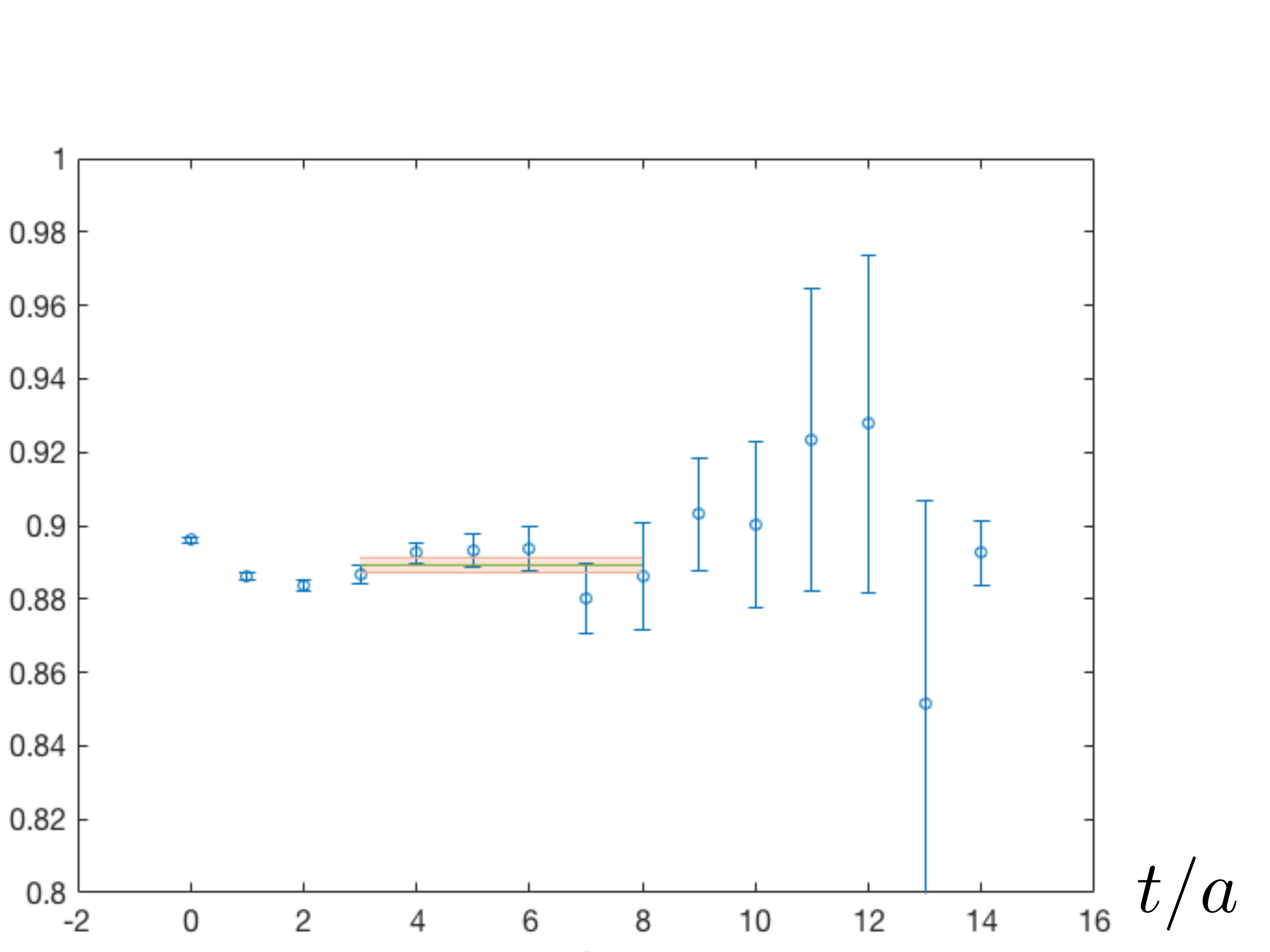}  \includegraphics[width=0.45\textwidth]{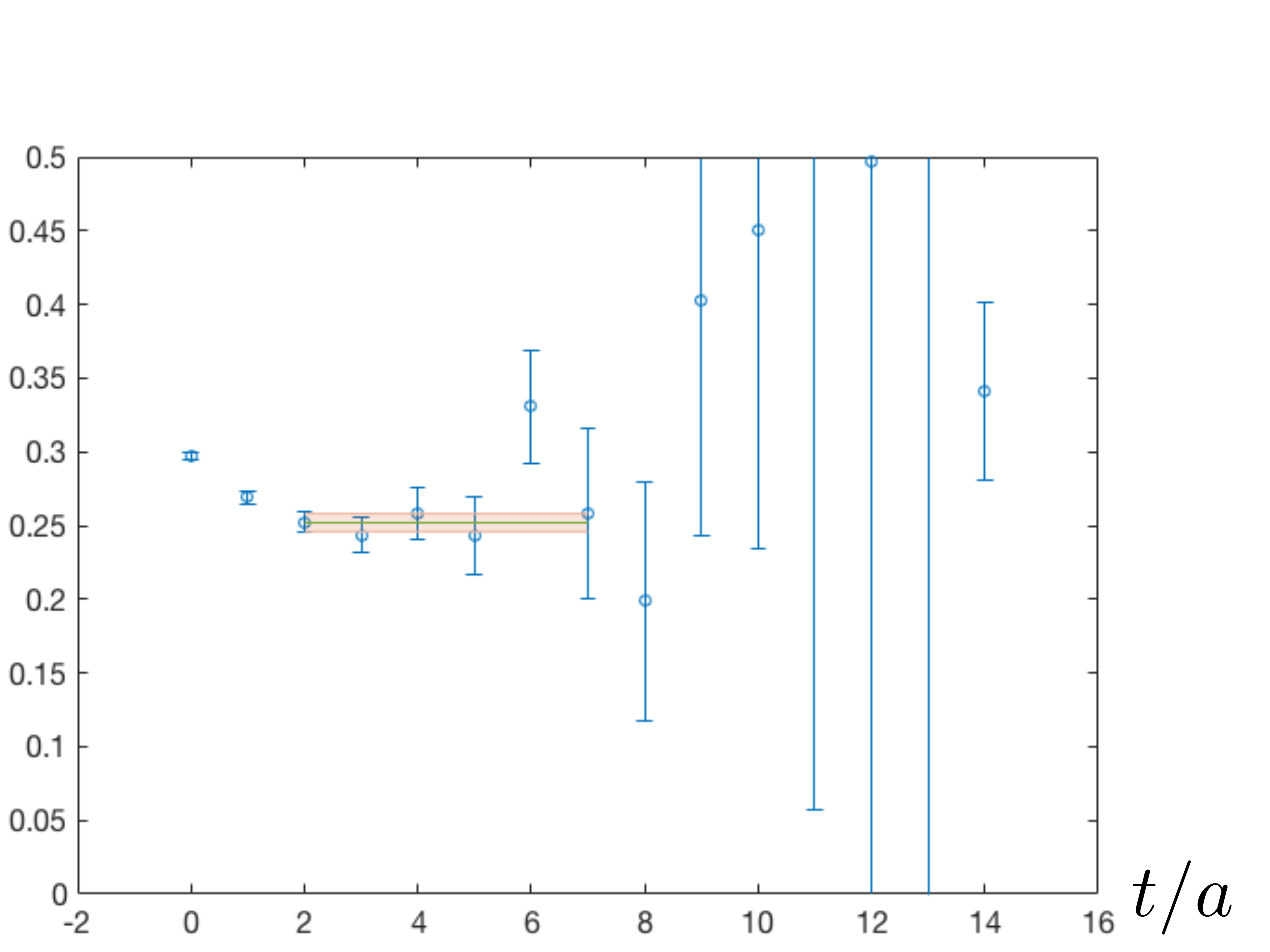}
  \caption{Typical fits used to extract the reduced matrix element. The upper panel corresponds to $p=2\pi/L \cdot 2$ and $z=4$ and the 
  lower panel to  $p=2\pi/L \cdot 3$ and $z=8$, where momentum and position are in lattice units.
  }
  \label{fig:plateau}
\end{figure}
This method of extracting the matrix element, contrary to the traditional sequential source approach, allows for the computation of the matrix element using all source-sink separations for the nucleon creation and annihilation operators. 

The  resulting effective matrix element has contaminations from excited states that scale as $e^{-t \Delta E }$,
where $t$ is the Euclidean time separation of the nucleon creation and annihilation operators,
and $\Delta E$ is the mass gap to the first excited state of the nucleon. Furthermore, it allows for the computation of  all nucleon matrix elements that correspond to different nucleon momentum spin polarization and flavor structure without additional computational cost. 

As a  result,  the total computational cost of this approach is less than the equivalent cost of performing the calculations with the sequential source method, especially because in our approach we put emphasis on having as many nucleon momentum states as possible.  This approach has recently been successfully used for both single and multi-nucleon matrix element calculations~\cite{Berkowitz:2017gql,Tiburzi:2017iux,Shanahan:2017bgi}.

In order to normalize our lattice matrix elements we note that, for $z_3=0$, the matrix element 
 $\mathcal{M}(z_3 P, z_3^2)$  corresponds to a  local vector (iso-vector) current, and therefore should  be equal to 1. However, on the lattice this is not the case due to lattice artifacts.
 Therefore we introduce a renormalization constant 
  \begin{equation}
  Z_P = \frac{1}{ \left.\mathcal{J}(z_3 P, z_3^2)\right|_{z_3=0}}\, .
  \label{eq:renorm}
  \end{equation}
 The factor  $Z_P$ has to be independent from $P$.  However,  again due to lattice
   artifacts or potential fitting systematics,  this is not the case. For this reason,  
   we renormalize  the matrix element for 
   each momentum  with its own $Z_P$ factor taking this way advantage of maximal statistical correlations {to reduce statistical errors},  as well as the cancellation of lattice artifacts in the ratio.
  Therefore,  our  matrix element is extracted  using the ratio
  \begin{equation}
\mathcal{M}(z_3 P, z_3^2)= \lim_{t\rightarrow\infty}  \frac{\mathcal{M}_{\rm eff}(z_3 P, z_3^2;t)}{\left.\mathcal{M}_{\rm eff}(z_3 P, z_3^2;t)\right|_{z_3=0}}  \  . 
  \end{equation} 
  In order to determine the reduced matrix element $\mathfrak{M}(\nu,z_3^2)$ we introduce the double ratio
  \begin{align}
\mathfrak{M}(\nu, z_3^2)=&  \lim_{t\rightarrow\infty} 
 \frac{\mathcal{M}_{\rm eff}(z_3 P, z_3^2;t)}{\left.\mathcal{M}_{\rm eff}(z_3 P, z_3^2;t)\right|_{z_3=0}} \nn 
 &\times 
\frac{\left.\mathcal{M}_{\rm eff}(z_3  P, z_3^2;t)\right|_{P=0,z_3=0}}
{\left.\mathcal{M}_{\rm eff}(z_3  P, z_3^2;t)\right|_{P=0}}  \ ,
\label{eq:redMatElem}
  \end{align} 
  which takes care of the renormalization of the vector current according to Eq.~(\ref{eq:renorm}).
  In practice, the   infinite $t$ limit is obtained with a fit to a constant  for  a suitable choice of a fitting range. In all cases we studied,  the average $\chi^2$ per degree of freedom was ${\cal O}(1)$. Typical fits used to extract the reduced matrix element are presented in Fig.~\ref{fig:plateau}. All fits are performed with the full covariance matrix and the error bars are determined with  the jackknife method.
  
  We note here that that the reduced  matrix element defined in Eq.~(\ref{eq:redMatElem}) has a well defined continuum limit and no additional renormalization is required. This continuum limit is obtained at fixed $\nu$ and $z^2$ as well as at fixed quark mass.

In this calculation we used momenta up to $6\cdot 2\pi/L $ along the $z$-axis. This corresponds to a physical momentum of about 2.5\,GeV.

\section{Discussion of  results}

\subsection{Rest-frame density and $Z$ factor}

\begin{figure}[h]
  \centerline{\includegraphics[width=3in]{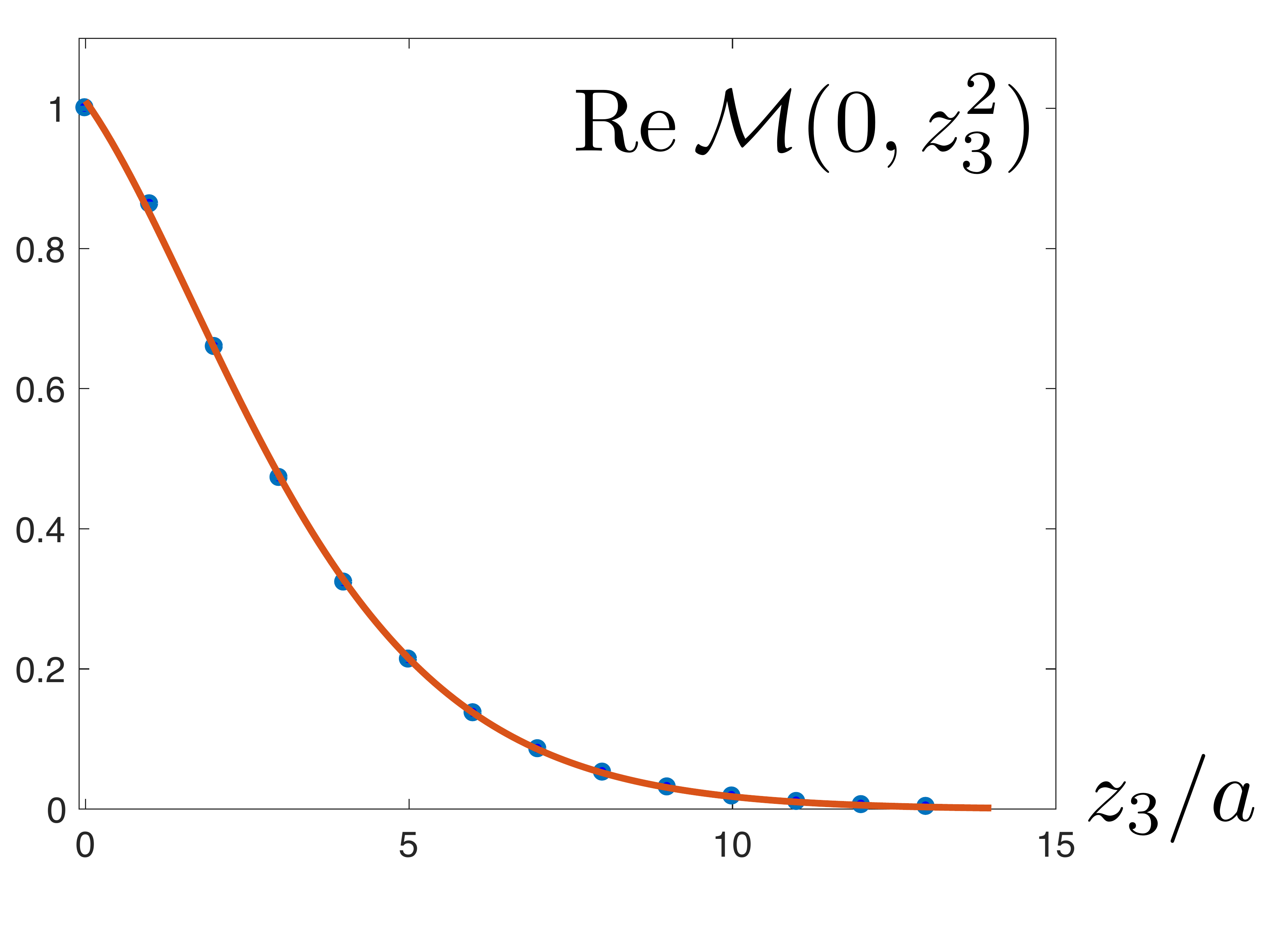}}
    \caption{Real part of the rest-frame  density  ${\cal M} (0,z_3^2)$ 
    \label{M0}}
    \end{figure}

An important object  is the rest-frame density 
${\cal M} (0,z_3^2)$. It is produced by data at $P=0$. 
The results for  its imaginary part are compatible with zero,
as required. 
The real part, shown in Fig. \ref{M0},   is a symmetric function of $z_3$,
and has a clearly visible linear component in its  fall-off with  $|z_3|$ for  small and 
middle values  of $|z_3|$.  In fact, a  linear exponential factor
$Z(z_3^2) \sim  e^{-c |z_3|/a}$ is expected as a manifestation
of the nonperturbative effects generated by  the straight-line gauge  link.

\subsection{Reduced Ioffe-time distributions}

\begin{figure}[t]
    \centerline{\includegraphics[width=3.5in]{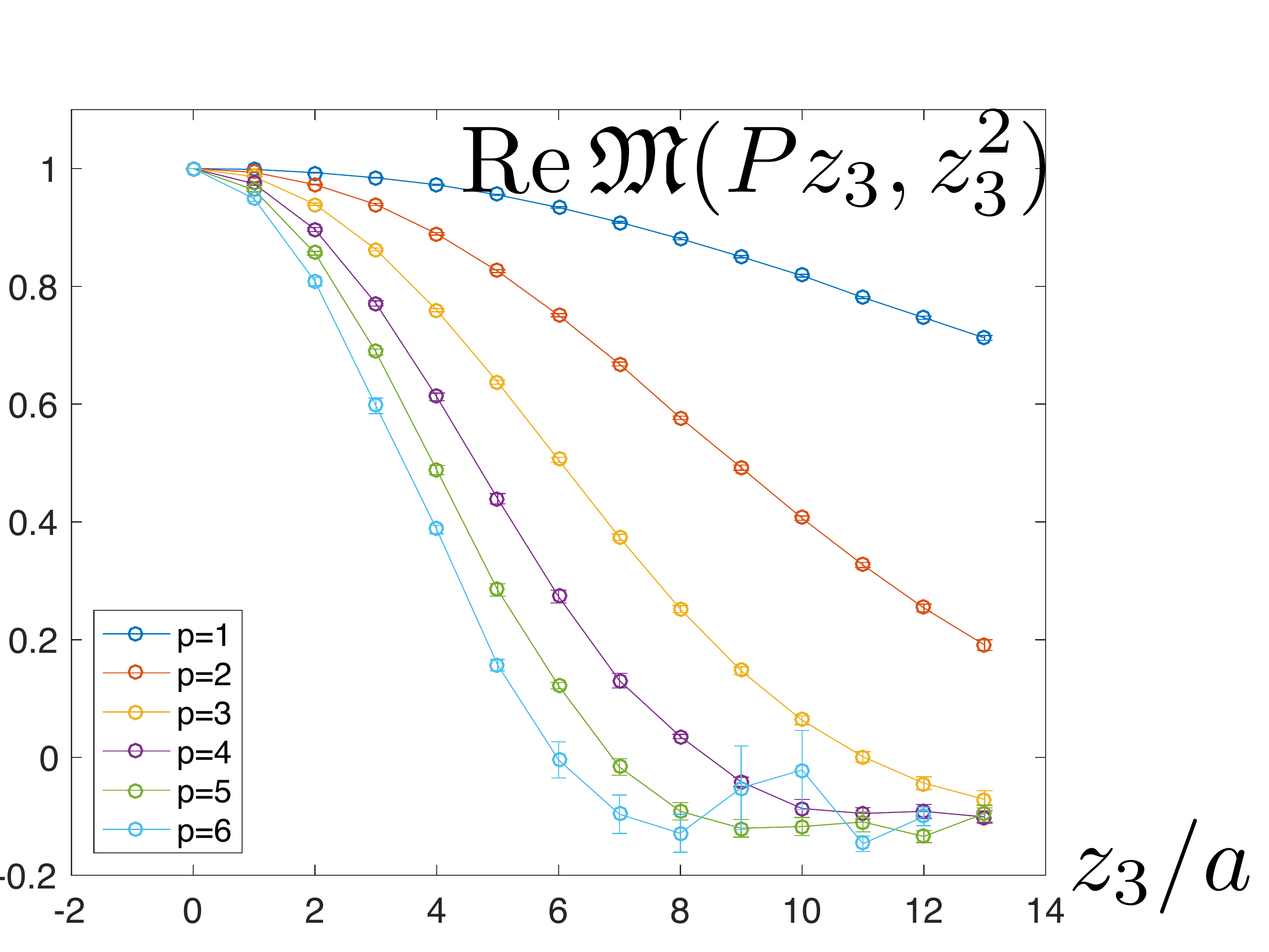} }
    \caption{Real  part of the reduced  distribution ${\mathfrak  M} (Pz_3,z_3^2)$ plotted as
    a  function of
    $z_3$. Here, $P= 2 \pi p/L$.
    \label{realz}}
    \end{figure}

 In   Fig. \ref{realz},  we plot the results for 
the real part of the  ratio  ${\cal M} (Pz_3, z_3^2)/{\cal M} (0,z_3^2)$ as a function of $z_3$ 
taken at six fixed values of the momentum $P$.  
One can see that  all the curves  have  a Gaussian-like shape.
Thus, the $Z(z_3^2)$  link renormalization factor has been canceled in the ratio,
as expected.

\begin{figure}[t]
    \centerline{\includegraphics[width=3.3in]{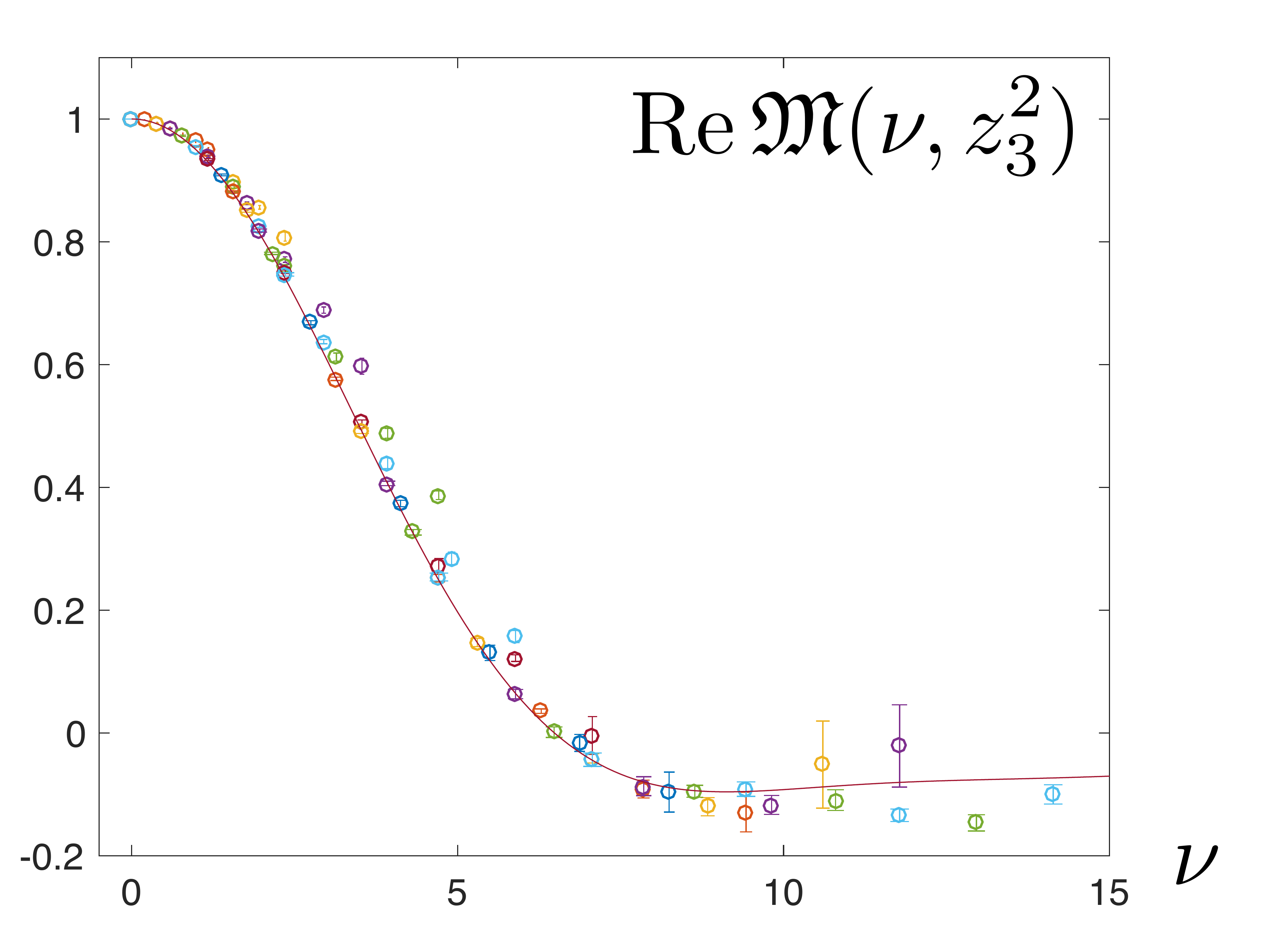}}
    \caption{Real  part of ${\mathfrak  M} (\nu, z_3^2)$  plotted as a function
    of $\nu =Pz_3$ and 
    compared to the curve given by Eqs. (\relax {\ref{MC}}), (\relax {\ref{qV}}).
        \label{realc}}
    \end{figure}

Furthermore, the curves  look  similar  to each other, {differing only by a decreasing  width  with $P$}. 
In  Fig. \ref{realc} , we plot   the  same data, but change the axis to $\nu=Pz_3$.
As one can see, now the  data practically fall on the same curve.
For  the  imaginary part, the situation is similar. 
 
 This phenomenon corresponds to factorization of the $x$- and $k_\perp$-dependence
    for the   soft TMD ${\cal F} (x, k_\perp^2)$, as discussed in previous sections.

      \begin{figure}[t]
    \centerline{\includegraphics[width=3.2in]{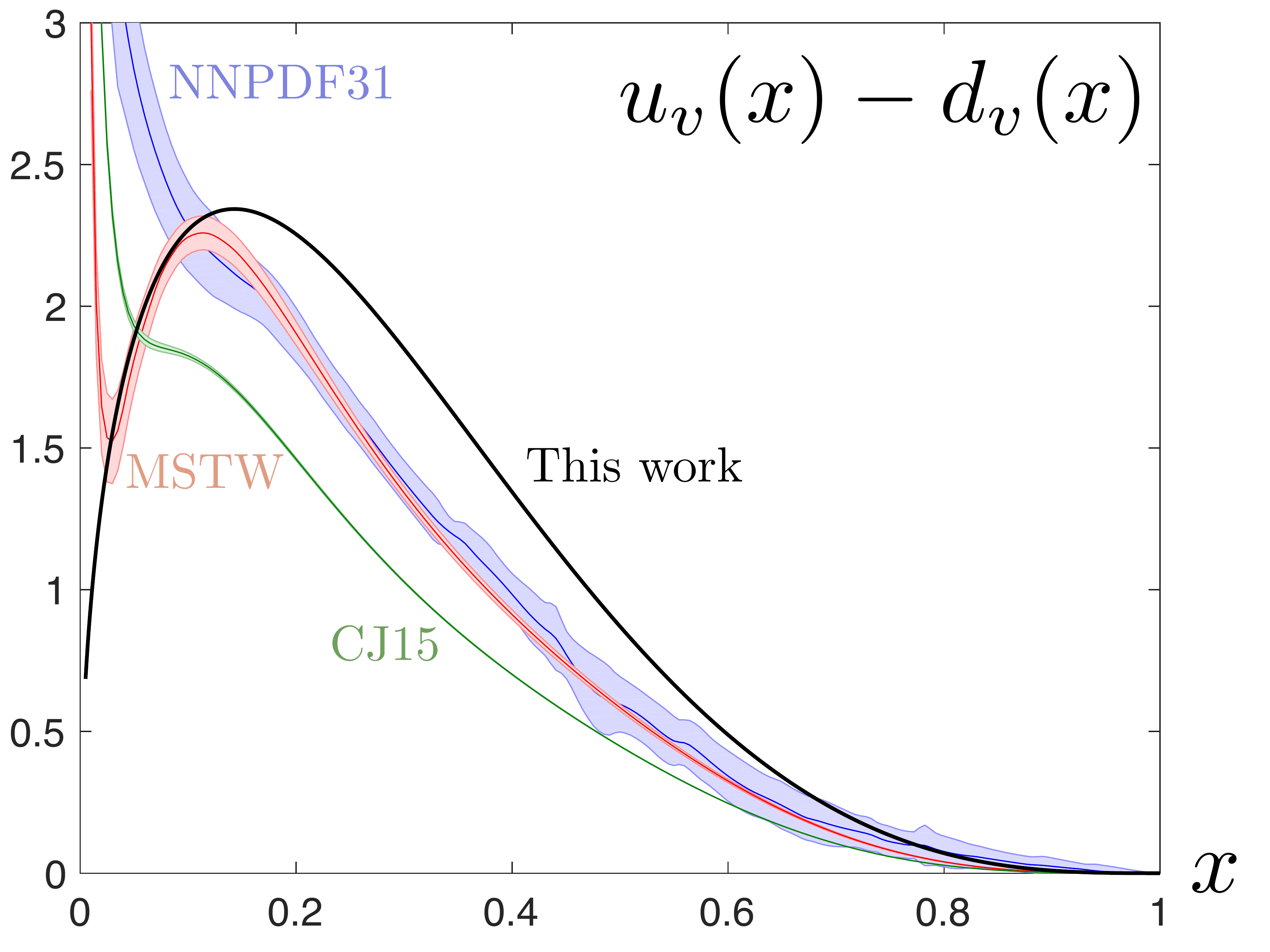} }
    \vspace{-0.35cm}
    \caption{Valence  distribution $q_v(x) $ as given  by \mbox{Eq. (\relax {\ref{qV}})} 
    compared with the  $Q^2=1$ GeV$^2$  NNLO global fits NNPDF31\_nnlo\_pch\_as\_0118\_mc\_164 \protect{\cite{Ball:2017nwa}  } and 
  MSTW2008nnlo68cl\_nf4   \protect{\cite{Martin:2009iq}}; and the NLO global fit CJ15nlo  \protect{\cite{Accardi:2016qay}},
  all extracted using the LHAPD6  library  \protect{\cite{Buckley:2014ana}}. 
  The bands around the global fits indicate their experimental and systematic uncertainties. 
        \label{MSTW}}
    \end{figure}

    \subsection{Quark-antiquark decomposition}
    
    The real part of the Ioffe-time distribution is obtained from  the cosine Fourier transform 
  \begin{align} 
{\mathfrak  M}_R (\nu) \equiv   & \int_0^1 dx \,  
\cos (\nu x)  \, q_v  (x)  
\label{MC}
     \end{align}
of the function  $q_v(x)$  given by the 
difference  $q_v(x) = q(x) -\bar q(x)$  of 
quark and antiquark   distributions. 
 In our case,   $q$ is  $u-d$ and 
    $\bar q = \bar u - \bar d$.
    The $x$-integral of $u-\bar u$ equals to the number of $u$-quarks 
    in the proton, which is 2, while
    the  $x$-integral of $d- \bar d$  equals 1. Thus, the $x$-integral of $q_v (x)$ should be equal to 1.

We found that our  data for the real part are well described  if one 
chooses       the function 
     \begin{align} 
    q_v(x) =  \frac{315}{32} \sqrt{x}  (1-x)^{3}\,  , 
    \label{qV}
    \end{align}
whose $x$-integral is normalized to 1.  
To get it, we formed cosine Fourier transforms ${\cal M} (\nu;a,b)$ of the 
normalized $x^a(1-x)^b$-type  functions 
and found the parameters $a,b$  by  fitting    our data. 
The comparison of the  data with the  curve based on
Eqs. (\relax {\ref{MC}}), (\relax {\ref{qV}})  is shown in Fig.  \ref{realc}. 

While all the data points were used in the fit, the latter  is clearly dominated by  
the points with the smaller values of   Re\ ${\mathfrak  M} (\nu, z_3^2)$.   
For $\nu <10$,  the data points  lying  above the curve, correspond to  
values of $z_3=3a$ to $5a$. As we will see later, they  reflect the perturbative evolution:  
\mbox{Re \ ${\mathfrak  M} (\nu, z_3)$}  increases when $z_3$ decreases.  
In this context, the overall  curve (\ref{qV}) corresponds to PDF  ``at low normalization point'', 
i.e.,  in  the region, where the perturbative evolution stops. 

In general,  it is more  appropriate to fit  Re  ${\mathfrak  M} (\nu, z_3^2)$ as a function of two variables,  
$\nu$ and $ z_3$, even though the dependence on $z_3$  is rather weak and  
noticeable  just for a few points. Since  we made a fit of  Re  ${\mathfrak  M} (\nu, z_3)$ 
as a function of just one variable $\nu$, there are points that visibly deviate 
from the curve, but we do not think that  
 it  makes    sense to translate the evolution $z_3$-dependence 
of small-$z_3$  points into    
an error band to our curve in \mbox{Fig.  \ref{realc}. }  
In the Section V.D, we evolve the data points to a common reference scale $z_0=2a$ 
and show the error band for the results obtained in this way.  

We realize   that our lattice setup  is rather crude (quenched approximation, very large pion mass), 
and for this reason  we do not attempt  to perform a thorough comparison of our results 
with experimental data. Still,  we think that some  kind of comparison is rather useful as  an illustration.  
 
Thus, we 
compare our $q_v (x) $ with three  global fits
for the difference \mbox{$u_v(x) -d_v(x)$}  of the valence distributions,  see Fig. \ref{MSTW}.
These global  fits curves correspond to   $\mu=1$ GeV scale, while our  
``low normalization point'' curve corresponds to $\mu \lesssim 0.3$ GeV.  
Still, one can see that our curve is not very far from the 
NNPDF31  \cite{Ball:2017nwa}  NNLO fit down  to $x=0.1$ 
 and from the   MSTW  \cite{Martin:2009iq} NNLO  fit  down  to $x=0.05$. 
 We also show the NLO fit CJ15  \cite{Accardi:2016qay}.

 Since the areas  under each  curve  are  equal to 1, our curve 
compensates the strong deficiency  in the $x<0.1$ region 
by exceeding the NNLO curves  at $x>0.1$ values.  
In other words, if our curve would better describe data in the $x<0.1$ region,
it would necessarily be smaller in the $x>0.1$ region.

       \begin{figure}[t]
    \centerline{\includegraphics[width=3.4in]{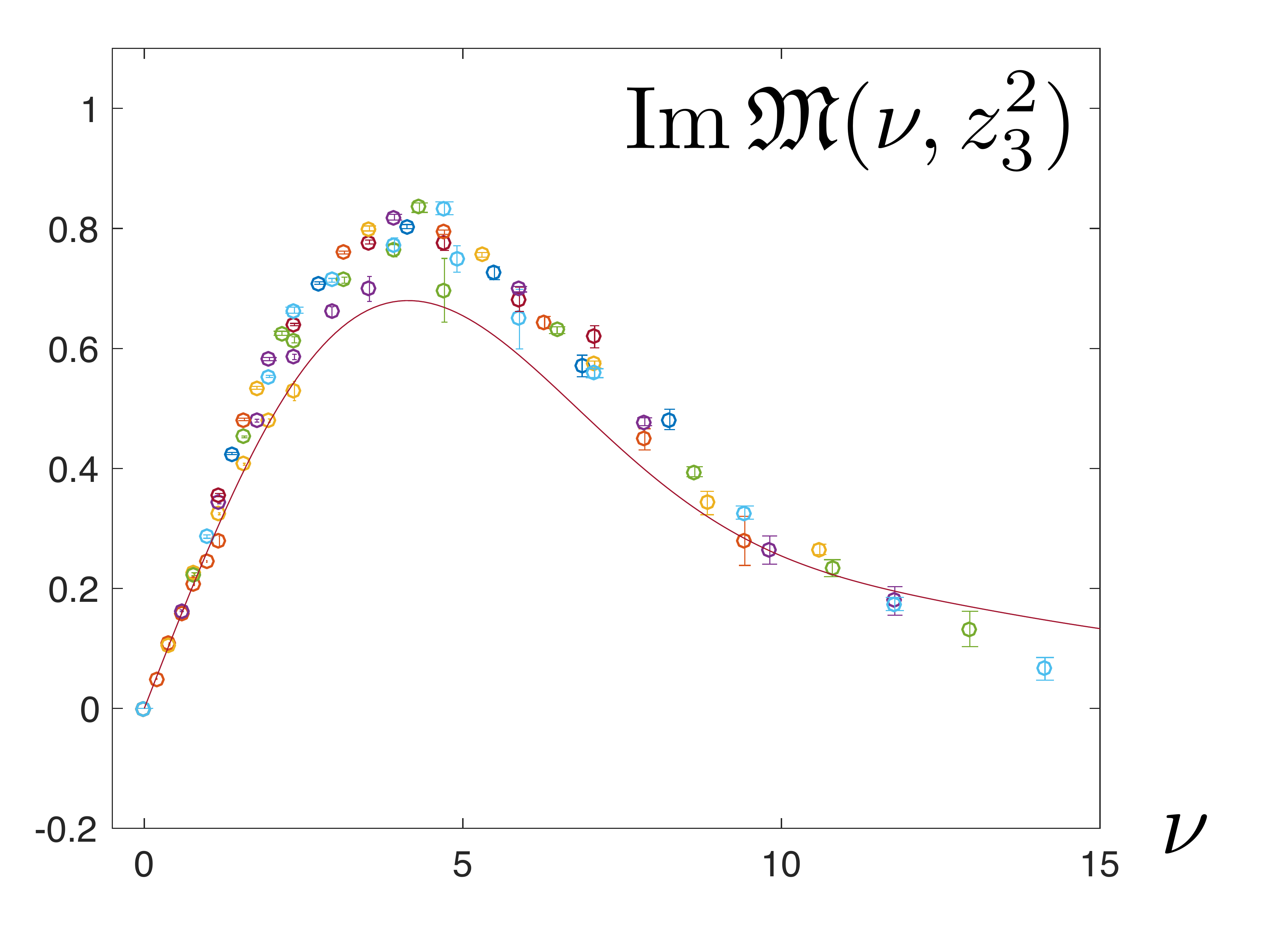} }
    \vspace{-0.2cm}
    \caption{Imaginary   part of ${\mathfrak  M} (\nu, z_3^2)$ 
    compared to the curve   $  {\mathfrak  M}_I^v (\nu)$  based on  $\bar q (x)=0$.
        \label{imc}}
    \end{figure}

      \begin{figure}[b]
    \centerline{ \includegraphics[width=3.4in]{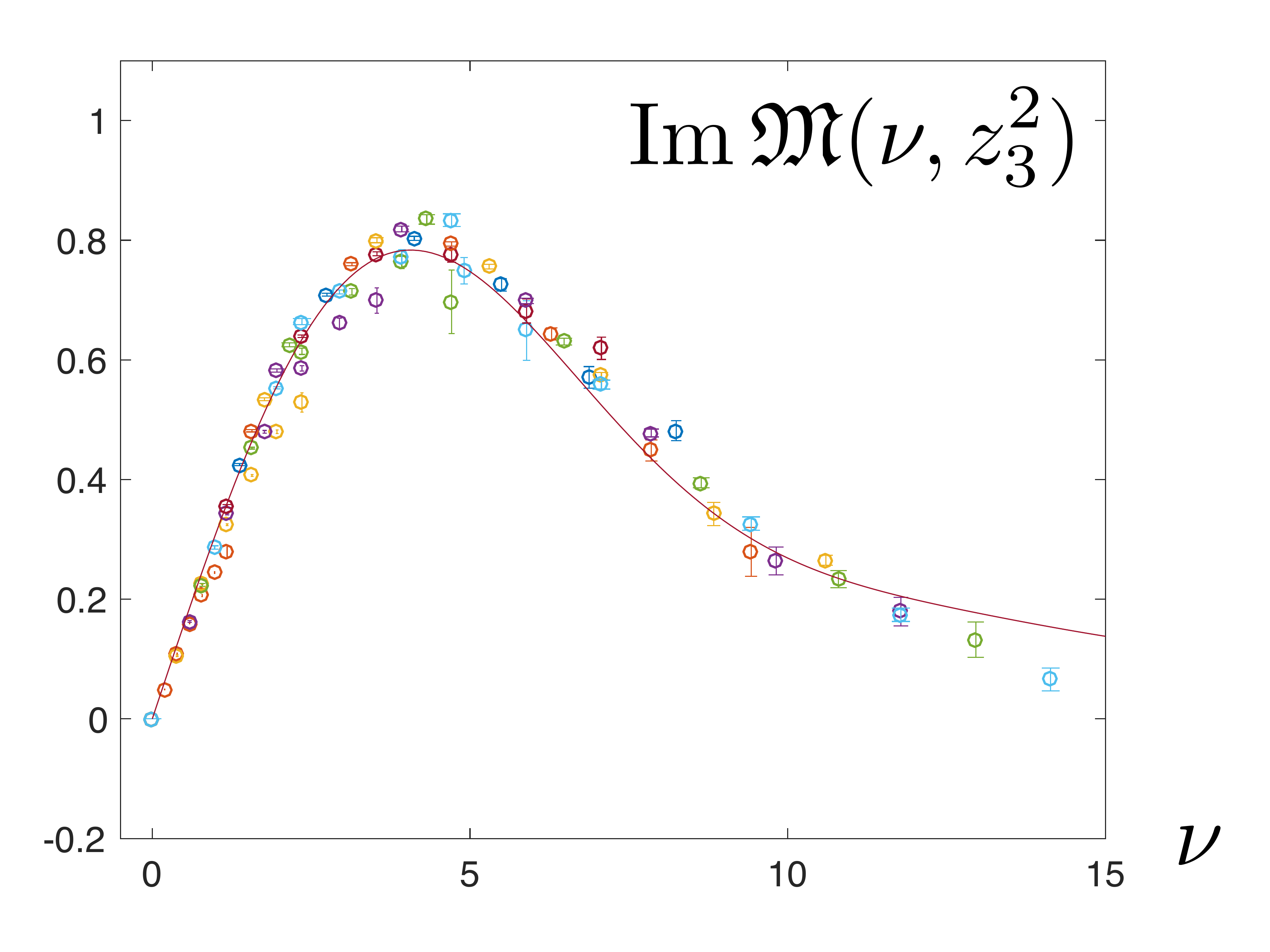}}
    \vspace{-0.2cm}
    \caption{Imaginary   part of ${\mathfrak  M} (\nu, z_3^2)$ 
    compared to the curve based on  $\bar q(x)$ given by Eq. (\relax {\ref{qA}}).
        \label{imca}}
    \end{figure}

\begin{figure}[t]
    \centerline{\includegraphics[width=3.5in]{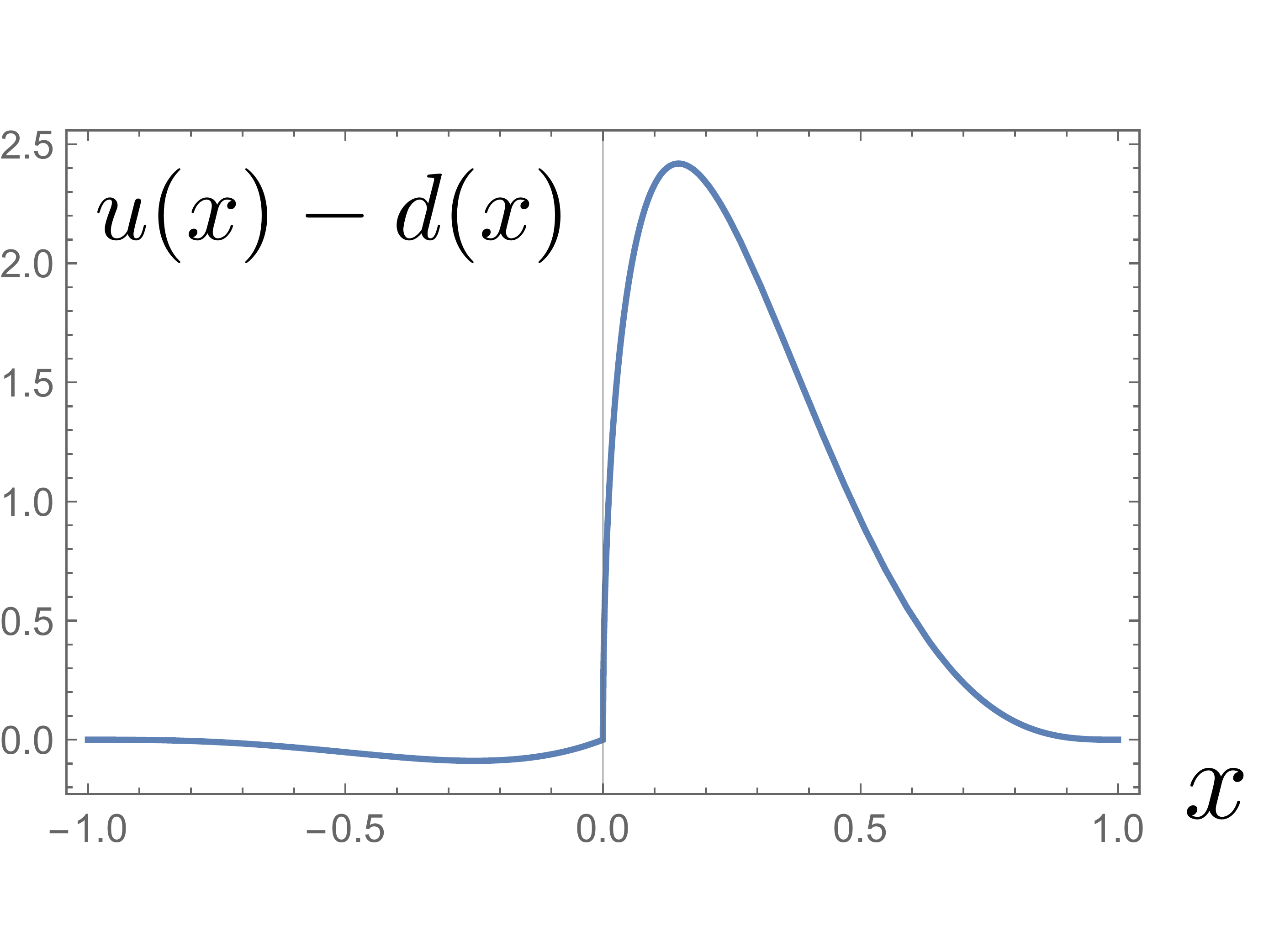} }
    \caption{Overall distribution $q(x) $ as defined by Eq. (\relax {\ref{q}}).  
        \label{umind}}
    \end{figure}

The sine Fourier transform 
      \begin{align} 
  {\mathfrak  M}_I (\nu)  \equiv  
& \int_0^1 dx \, 
\sin (\nu x) \, q_+  (x) 
     \end{align} 
   is  built from  the function $q_+ (x)= q(x) + \bar q (x)$,
   which may be also represented as \mbox{$q_+ (x)= q_v(x) + 2 \bar q (x)$. }
   If we neglect the antiquark contribution and use \mbox{$q_+  (x) = q_v(x)$,}
   we get the curve shown in   Fig. \ref{imc} 
    (call   it $  {\mathfrak  M}_I^v (\nu)$). The agreement with the data
   is strongly improved if we use a non-vanishing antiquark contribution, namely
        \begin{align} 
    \bar q(x) =\bar u (x) - \bar d(x) =  0.07 \left [ 20  \,  x (1-x)^{3} \right ] \,  ,
    \label{qA}
    \end{align} 
    see Fig. \ref{imca}.  
    This function was obtained by fitting the data for 
    the difference ${\rm Im} \ {\mathfrak M} (\nu, z_3^2) - {\mathfrak M}_I^v(\nu)$ 
    by sine Fourier transforms of $A x^a (1-x)^b$ functions. 
     This result corresponds to  
          \begin{align} 
 \int_0^1 dx \, [\bar u (x) - \bar d(x)] =  0.07 \ . 
     \end{align}
     The combined distribution 
         \begin{align} 
q(x)& = u(x)-d(x)    \nn & =  [q_v (x) +\bar q (x)  ]\, \theta(x>0) 
-
\bar q (-x)  \, \theta(x<0) 
\label{q}
     \end{align}  
 defined on the $-1\leq x \leq 1$ interval
     is shown in Fig. \ref{umind}.

        \subsection{Evolution}

    While an overall agreement of the data
    with a \mbox{$z_3$-independent} curve  looks satisfactory,
    one can easily notice  a residual \mbox{$z_3$-dependence}  in the data.  
    It  is  especially visible when, for a particular $\nu$,    there  are several  data points 
  corresponding to different values of $z_3$.
It is interesting to check if this dependence corresponds to perturbative evolution.

To begin with,  the  evolution of the real part should  lead
to its  {\it decrease} when $z_3^2$ increases. On the other hand, as pointed out at the end of section III, 
 the function  ${{\rm Im} \, \cal M} ( \nu, z_3^2)$
{\it increases} 
 when $z_3^2$ {\it increases}  as long as  \mbox{$\nu \lesssim 5.5$.}
 Our data follow these patterns.

As we discussed, the evolution corresponds to $\ln z_3^2$  singularities of the Ioffe-time distributions
for small $z_3^2$.  Thus, a natural idea is  to check if the data  corresponding to 
small  $z'_3$ and $z_3$  may be related by  
  \begin{align}
{\mathfrak M} (\nu, {z'}_3^2) {=}
{\mathfrak M} (\nu, z_3^2) \
 -\frac23  \frac{\alpha_s}{\pi}   \ln ({z'_3}^2/z_3^2) B \otimes {\mathfrak M}\,  (\nu, z_3^2)   
 \label{Prop}
 \end{align}
 for some value of $\alpha_s$. 
Here $B$ is the evolution kernel  (\ref{Bu}). 
In our case,
    \begin{align}
B \otimes {\mathfrak M}\,  (\nu)   & =
\int_0^1  du \,  \frac{1+u^2}{1- u} \,    [{\mathfrak M} ( \nu)- {\mathfrak  M} (u \nu)  ]
 . 
\label{plusM}
 \end{align}

 More specifically, we fix  the point  $z'_3$ at the   value \mbox{$z_0=2a$}   corresponding, 
 at the leading logarithm level,  
 to the
 $\overline {\rm MS}$-scheme scale \mbox{$\mu_0 = 1$  GeV }  and build the function 
   \begin{align}
\widetilde{\mathfrak M} (\nu, z_0^2) {\equiv }
{\mathfrak M} (\nu, z_3^2) \
 -\frac23  \frac{\alpha_s}{\pi}   \ln (z_0^2/z_3^2) B \otimes {\mathfrak M}\,  (\nu, z_3^2)  
 \label{EvoM}
 \end{align}
 from the data points for ${\mathfrak M}\,  (\nu, z_3^2)$
 using various values for $\alpha_s$.   
 
 Since the perturbative evolution is expected for small $z_3$, we include in this analysis
 the data with $z_3$ up to 4
lattice spacings, which corresponds to energy scales \mbox{$\mu = 2,     1,     0.7$  and     0.5  GeV.}

                 \begin{figure}[t]
    \centerline{ \includegraphics[width=3.4in]{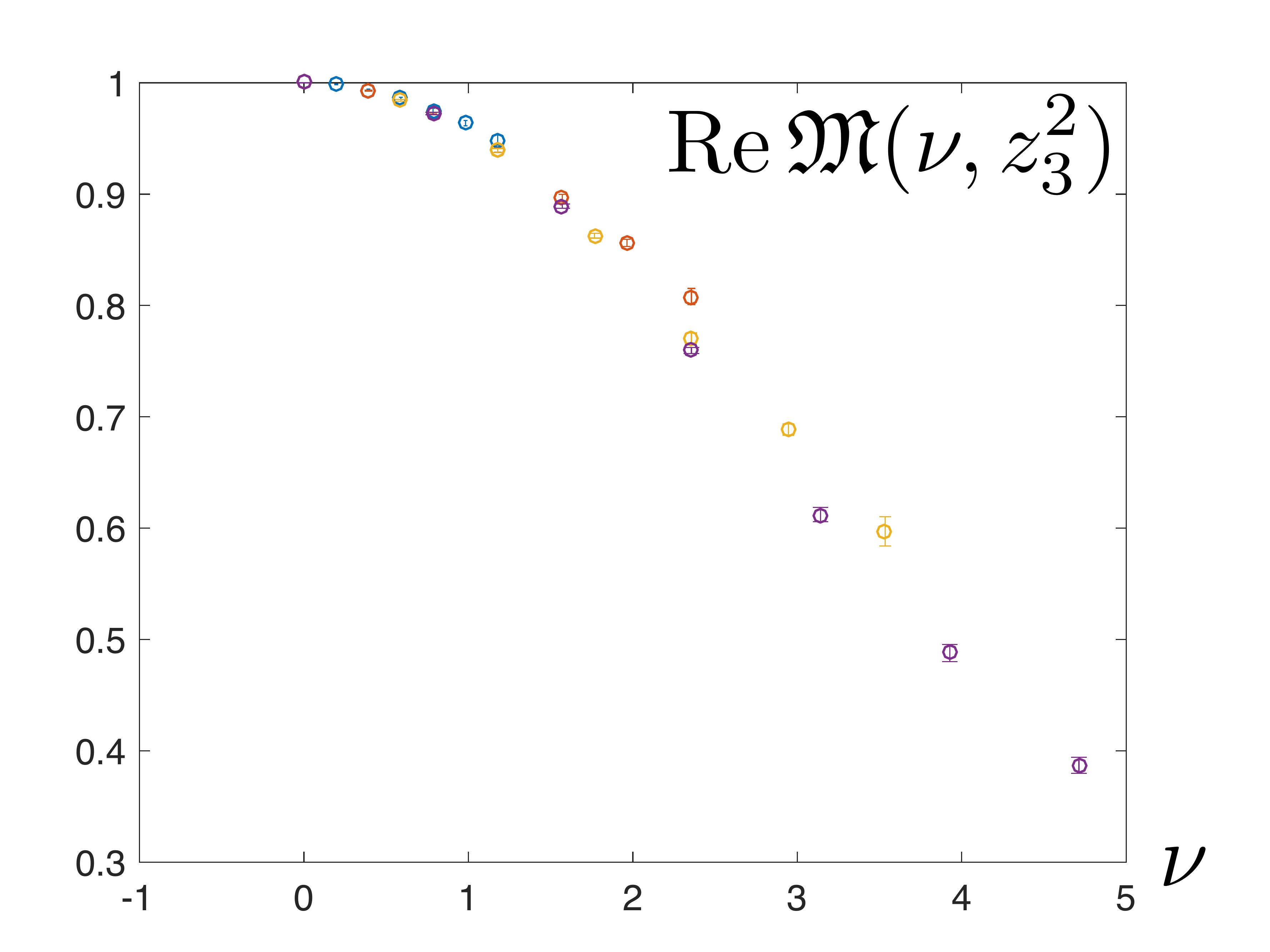}}
    \caption{Real    part of ${\mathfrak  M} (\nu, z_3^2)$ 
    for $z_3/a =1, 2, 3,$ and 4.
        \label{ER0}}
    \end{figure}

      \begin{figure}[t]
    \centerline{ \includegraphics[width=3.3in]{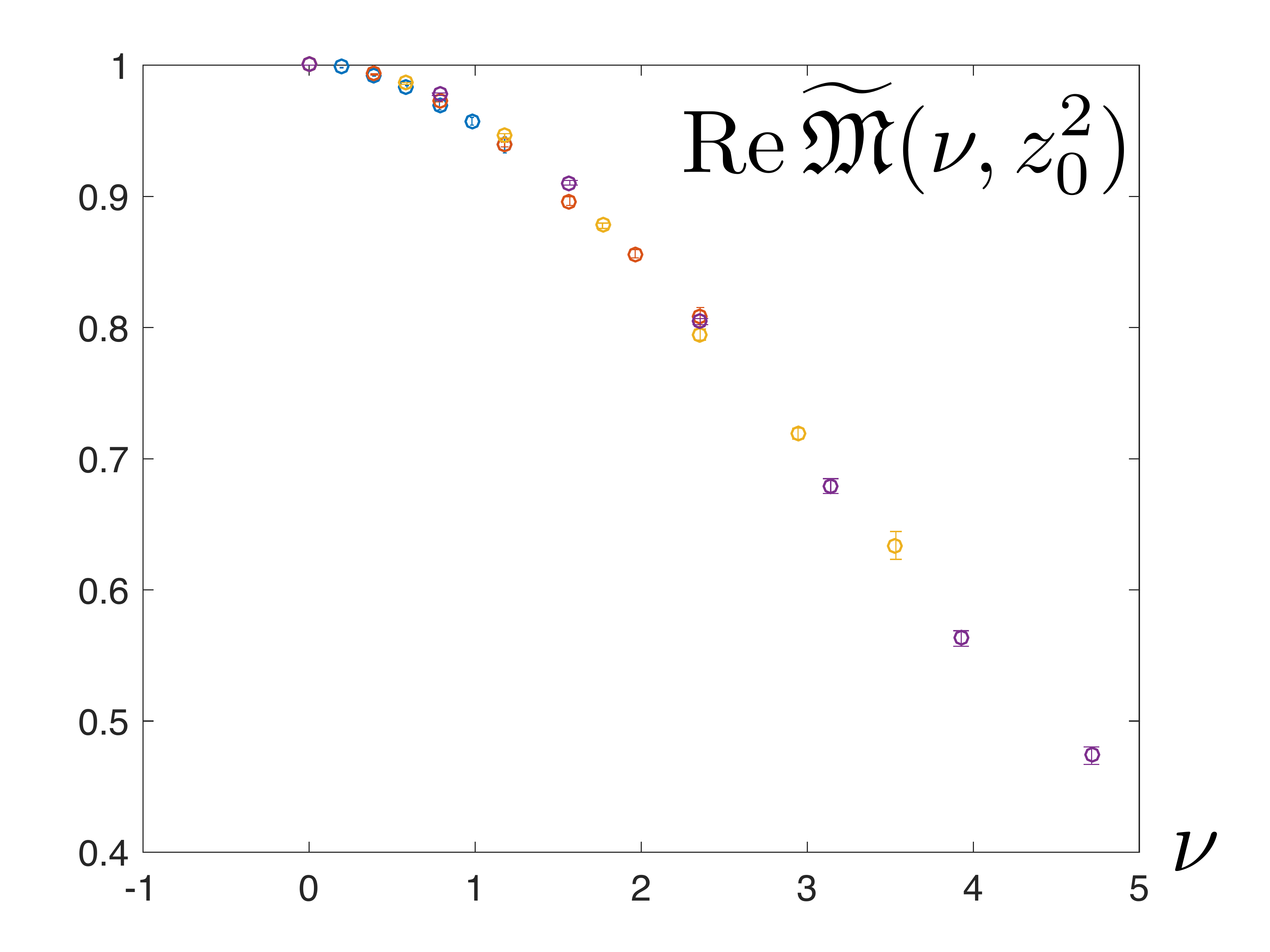}}
    \caption{Evolved data points for the real part.
        \label{ERE}}
    \end{figure}

For the real part, these data points are shown in  \mbox{Fig. \ref{ER0}. }
 As one can see, there is a visible scatter of the data points.  
    Using $\alpha_s/\pi =0.1$, we calculate the
    ``evolved'' data points corresponding to  the function $\widetilde{\mathfrak M} (\nu, z_0^2) $.
    The results are shown in Fig. \ref{ERE}.  The evolved data points  are now very close
    to a universal  curve.

    \begin{figure}[t]
    \vspace{0.3cm}
    \centerline{ \includegraphics[width=3.3in]{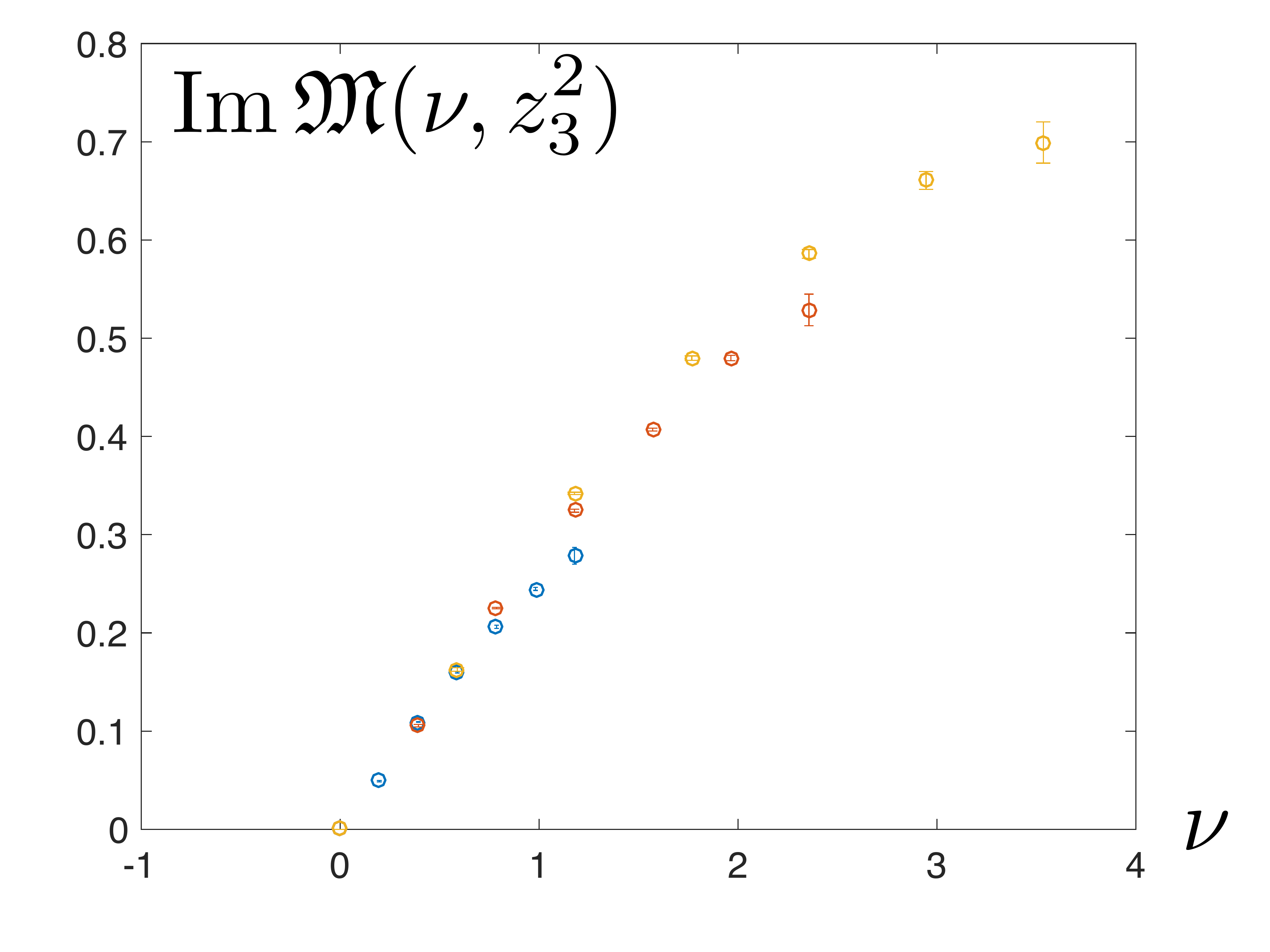}}
    \caption{Imaginary     part of ${\mathfrak  M} (\nu, z_3^2)$ 
    for $z_3/a =1, 2, 3,$ and 4.
        \label{EI0}}
    \end{figure}
    
           \begin{figure}[t]
    \centerline{ \includegraphics[width=3.3in]{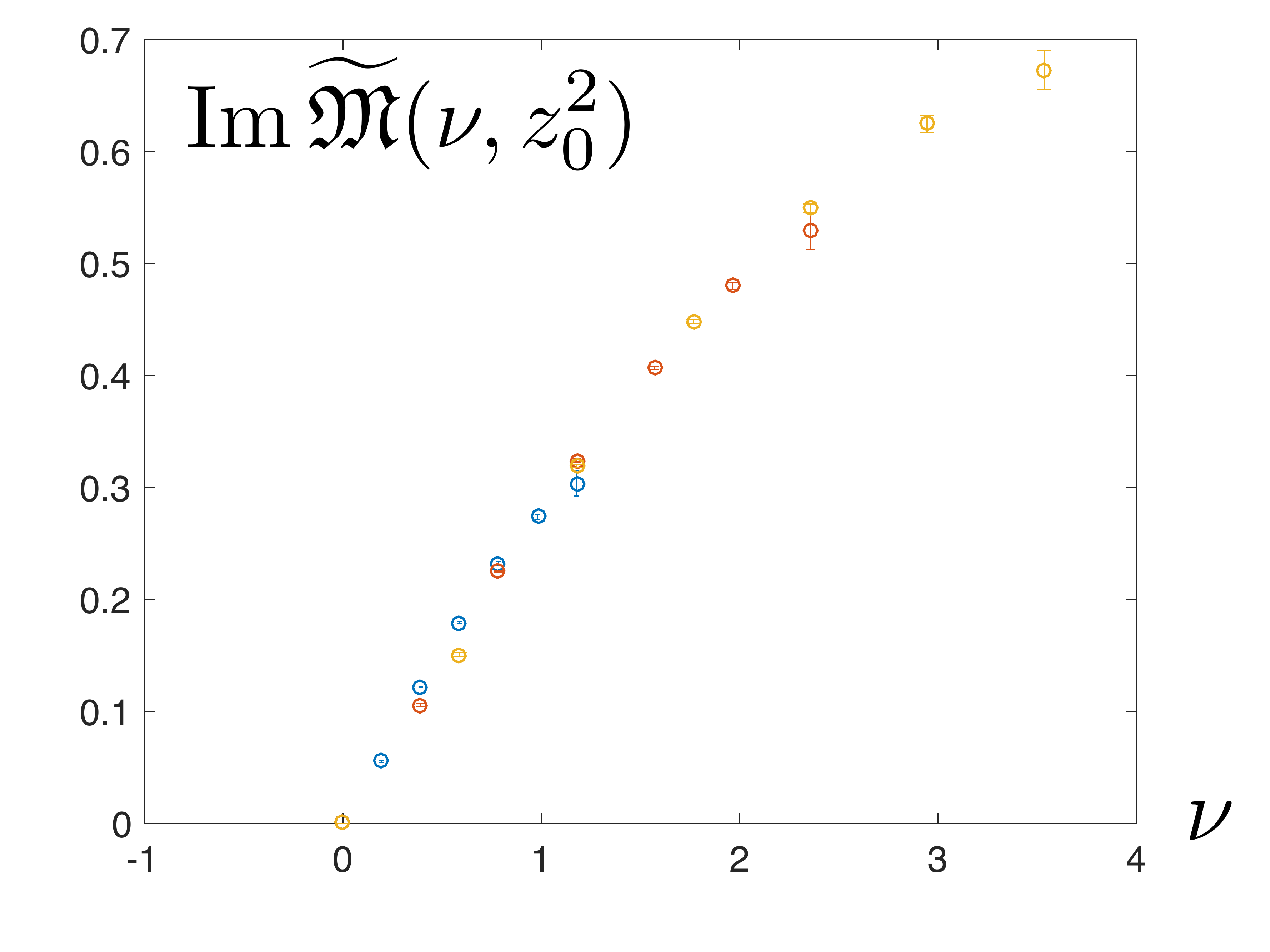}}
    \vspace{-0.3cm}
    \caption{Evolved data points for the imaginary part.
        \label{EIE}}
    \end{figure}

        \begin{figure}[t]
    \centerline{ \includegraphics[width=3.5in]{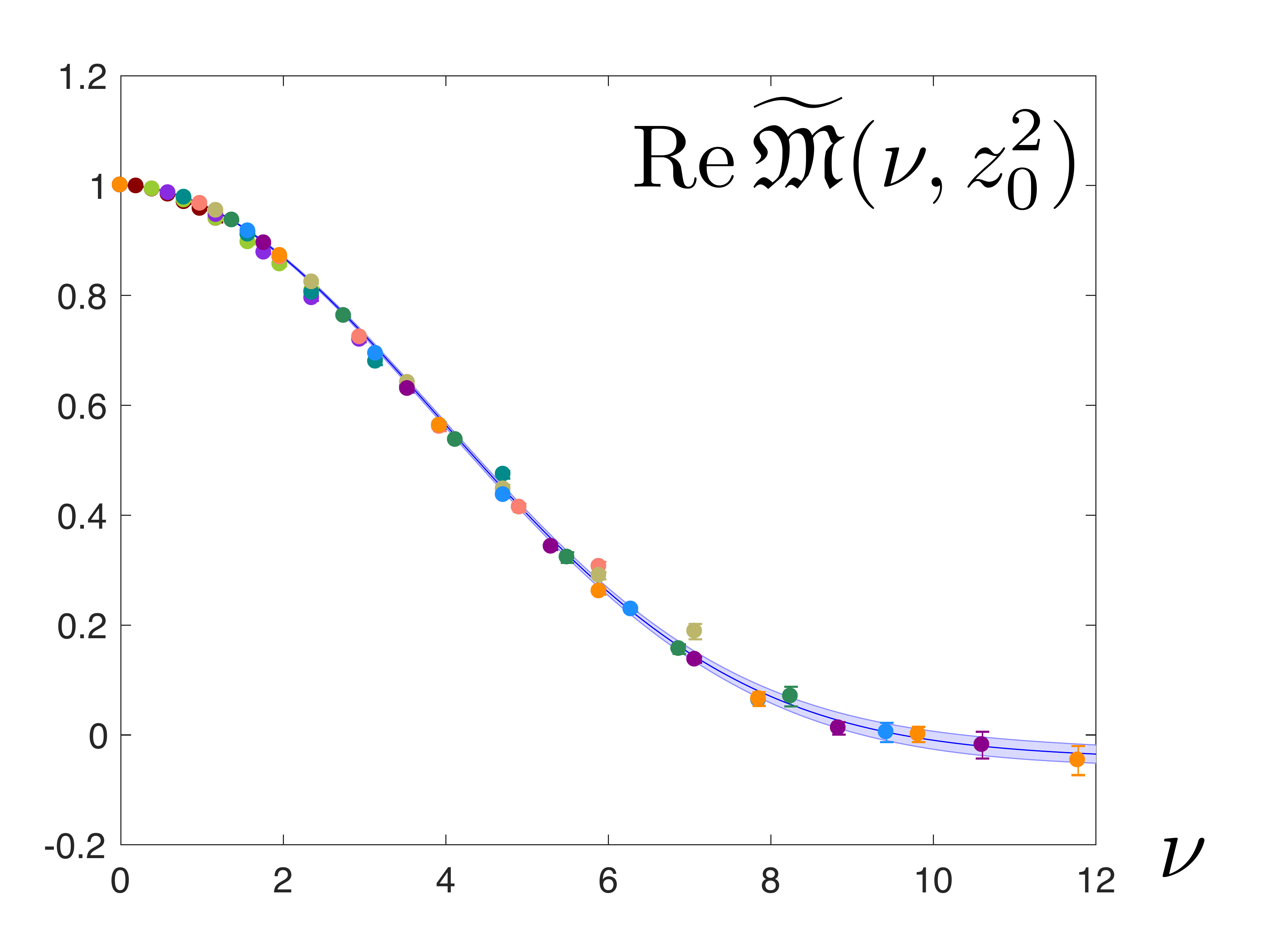}}
    \vspace{-0.3cm}
    \caption{Data points for Re ${\mathfrak M}\,  (\nu, z_3^2)$ with $z_3 \leq 10a$ 
     evolved to $z_0=2a$ as described in the text.
        \label{evpoi}}
    \end{figure}

        \begin{figure}[b]
   { \includegraphics[width=4in]{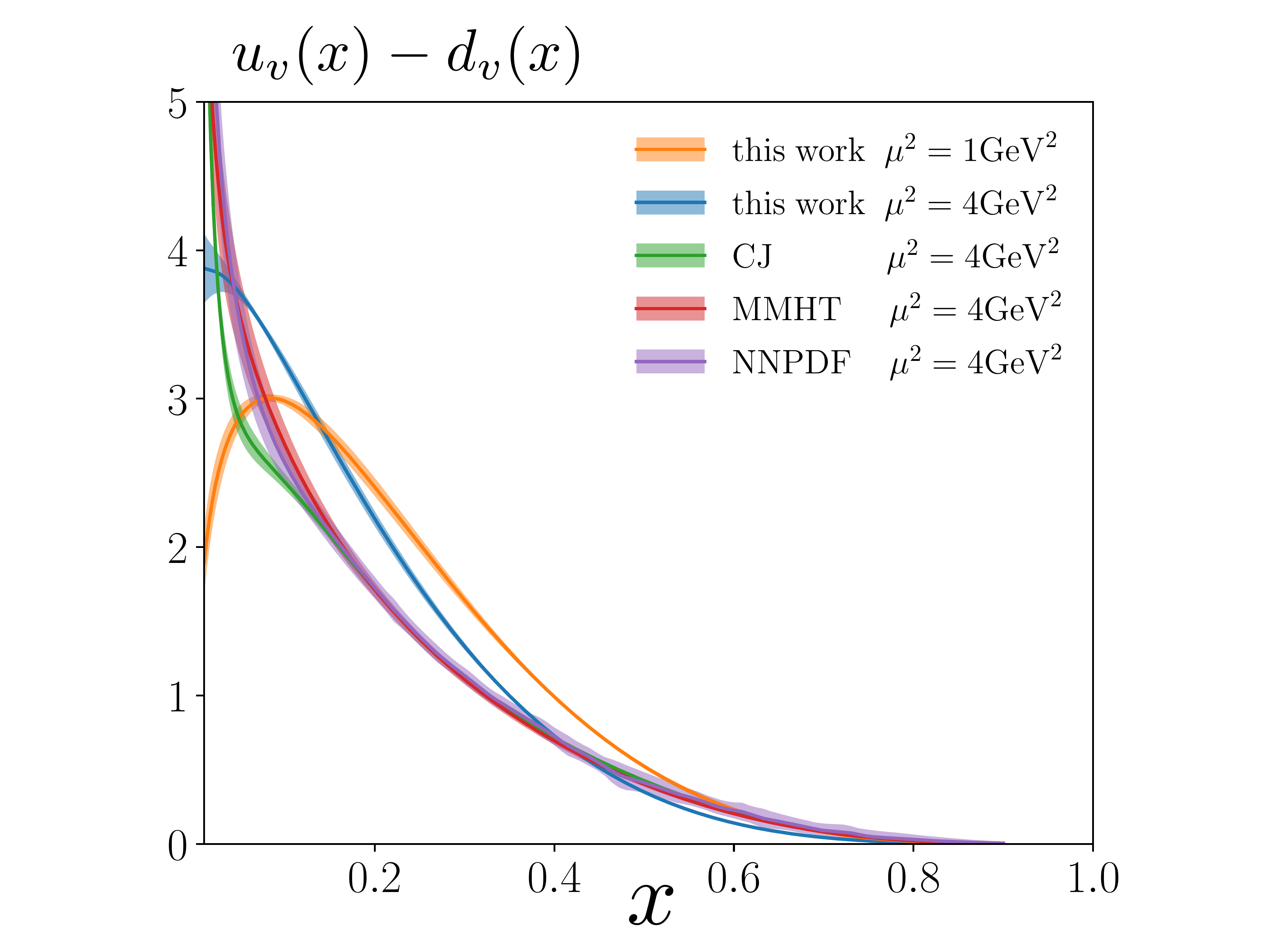}}
    \vspace{-0.3cm}
    \caption{
    Curve for $u_v(x) - d_v(x)$ built from the evolved data
   shown in Fig. \ref{evpoi}, and     treated as corresponding to
     the  \mbox{$\mu^2 =1$  GeV$^2$}   scale; 
then  evolved to the reference point  $\mu^2 =4$  GeV$^2$  of the  global fits.
        \label{evolved}}
    \end{figure}

  In  Fig. \ref{EI0},    we show the initial data points for the imaginary  part. The evolved data points
  constructed using the same  $\alpha_s/\pi =0.1$ value  are shown  in Fig. \ref{EIE}. 
         Again, they are close to a universal curve.  
         This analysis indicates that the residual $z_3^2$-dependence 
          of ${\mathfrak M}\,  (\nu, z_3^2)$ at fixed $\nu$ is compatible with the expected logarithmic evolution at small $z_3^2$.
 Clearly this is an important feature of our calculation which needs to be further studied as it will play an essential role in reliable extraction of renormalized PDFs from this type of lattice calculations.

With a smaller lattice spacing, the use of perturbative evolution may be justified  
in a wider region of $\nu$. While our data extend to rather large 
separations $\sim 1$ fm, we find it instructive  to use them as an example to   
 illustrate the trends generated  by 
the perturbative evolution.    

To this end, we applied the leading logarithm formula 
(\ref{EvoM}) with $z_0=2a$ and $\alpha_s/\pi=0.1$ to our data points with $z_3 \leq 6a$.  
Assuming that evolution stops for $z_3 \gtrsim 6a$ (as indicated by our data), 
the data points with \mbox{$7a \leq z_3 \leq 10a$}   
 were evolved to $z_0$ using Eq. (\ref{EvoM}) with $z_3=6a$. 
The  data points evolved in this way are shown in Fig. \ref{evpoi}.    

Fitting  the evolved  points by  cosine Fourier transforms ${\cal M} (\nu;a,b)$ of the 
normalized $N(a,b) x^a(1-x)^b$-type  functions,  
we  found that they may be described if one takes $a = 0.36(6)$  and $b=3.95(22)$.  
Treating $z_0=2a$ as the $\overline{\rm MS}$ scale $\mu=1$ GeV, one can further evolve  
the curve to the standard reference scale $\mu^2 = 4$ GeV$^2$ of the global fits, see Fig. \ref{evolved}.    
Comparing with Fig. \ref{MSTW}, we see that the perturbative evolution shifts   
our curves,  moving them  closer to the global fits.   


   \section{Summary}  
   
   In this paper, we demonstrated   a new method of extracting
  parton distributions from lattice calculations.  
   It is  based on the  ideas, formulated in \mbox{Ref. \cite{Radyushkin:2017cyf}.}
   
   First, we   treat the generic equal-time  matrix element as a function ${\cal M} (\nu, z_3^2)$ 
   of the Ioffe time $\nu = Pz_3$ and the distance $z_3$. The 
   next idea is to  form the ratio ${\mathfrak  M} (\nu, z_3^2) \equiv {\cal M} (\nu, z_3^2)/{\cal M} (0, z_3^2)$
   of the Ioffe-time distribution ${\cal M} (\nu, z_3^2)$ and the rest-frame density given by 
    ${\cal M} (0, z_3^2)$.

    Our lattice calculation clearly shows the presence of a linear  component in the $z_3$-dependence 
    of  the rest-frame function, that may be attributed to the expected $Z(z_3^2) \sim e^{-c|z_3|/a}$ behavior 
    generated by the gauge link.
    On the next step, we observe that  the ratio ${\cal M} (Pz_3 , z_3^2)/{\cal M} (0, z_3^2)$
    has a Gaussian-type  behavior with respect to $z_3$ for all 6 
    values of $P$ that were used in the calculation. 
    This means that $Z(z_3^2)$ factors entering into the numerator and denominator
    of the ${\mathfrak  M} (Pz_3, z_3^2)$  ratio have been canceled, as expected.
    
    Still, there is no {\it a priori} principle predicting that  the  remaining 
    non-logarithmic  $z_3^2$-dependence cancels 
    between  the numerator and the denominator of the ratio ${\cal M} (\nu, z_3^2)/{\cal M} (0, z_3^2)$.
    Such a   $z_3^2$-dependence can be removed if needed with a systematic fitting procedure 
    from which the Ioffe time PDF will be  extracted in the $z_3^2=0$ limit.

    However, we found  that when plotted as a function of $\nu$ and $z_3$, the data both 
    for the real and imaginary parts of ${\mathfrak  M} (\nu, z_3^2)$  are very close 
    to  the respective universal functions. This observation indicates  that 
    the soft part of the \mbox{$z_3^2$-dependence}  of $   {\cal M} (\nu, z_3^2)$ 
    has been  canceled by the rest-frame density $   {\cal M} (0, z_3^2)$.
    This phenomenon corresponds to factorization of the $x$- and $k_\perp$-dependence
    for the  TMD ${\cal F} (x, k_\perp^2)$. 
    
    While this evidence in favor of the factorization property is an important result on its own,
    we want to stress  that our approach  {\it is not based}
    on the factorization.  It is based on the use of the ratio 
    ${\cal M} (\nu, z_3^2)/{\cal M} (0, z_3^2)$. Its   residual soft $z_3^2$-dependence 
    may be systematically analyzed and fitted, so that the $z_3^2$-limit 
    may be taken in a controllable way.  
    
    Luckily, the  data do  not show a visible 
  polynomial dependence on $z_3^2$  
  within our current statistical and systematic errors. In future work we intend to carefully study 
  the residual polynomial  $z_3^2$ 
  effects  and incorporate them in the extraction of PDFs using the lattice methodology introduced here.
    
    In addition,  
    we have checked that, for small $z_3 \leq 4a$,  the residual  $z_3$-dependence
    may be explained by  perturbative evolution, with the $\alpha_s$ value
     corresponding to  $\alpha_s/\pi =0.1$. We have evolved these small-$z_3$ 
     data points to the $z_3=2a$ scale, which corresponds to $\mu^2=1$ GeV$^2$.
     The evolved data better approximate  universal curves both for real 
     and imaginary parts of ${\cal M}$, supporting the argument that perturbative evolution is observed.

Thus, these  $ \nu \lesssim 4$ parts  of the     universal curves may be treated 
as corresponding to the $\mu=1$ GeV  scale.  
 Other  data points correspond 
 to  $z_3  >4a$ values, and   formally should  be treated 
 as corresponding to scales $\mu \lesssim 0.3$ GeV.
 All these data points basically lie on the 
same universal curve.  This indicates that evolution stops at such scales. 
We  compared  this  ``low normalization point''  curve 
with three global fits evolved to the $\mu =1$ GeV scale,
and observed that  our curve (\ref{qV}) for the valence  $u_v(x)- d_v (x)$ 
 distribution  shows 
     the $(1-x)^3$ behavior for $x \to 1$  in accord with usual expectations.   
     Also, it rather closely follows the NNPDF31  and, 
 especially,   MSTW NNLO global fits down to rather  small  $x$ values.

     Still,   our curve  strongly deviates from the global fits   for \mbox{$x <0.1$  } in the NNPDF31  case 
and for  \mbox{$x <0.05$  } in the MSTW case. 
However, the shape of PDFs is  affected  by the perturbative evolution.  
To illustrate the scope of these effects, we evolved all our points with $z_3 \leq 10a$   
to a universal scale $z_0=2a$ corresponding to $\mu =1 $ GeV, 
and then further evolved the resulting PDF to  $\mu =2 $ GeV, 
that is the standard reference scale for global fits. 
Our final curve is rather close to these fits,
which demonstrates that the perturbative evolution 
plays an important role in comparison of lattice results with the data.
Again, one needs smaller lattice spacings to justify the use 
of the perturbative evolution equation in a sufficiently wide 
interval of Ioffe time parameters $\nu$.

     The data also indicate a nonzero {\it positive}  antiquark distribution $\bar q (x) = \bar u(x) - \bar d(x)$.
     It changes the $x$-integral of $q(x)$ by 7\% and has $\sim x (1-x)^3$ behavior. 
     Since we are using the quenched approximation, these antiquarks come from 
     ``connected diagrams''. Hence,  one should expect that the ratio $\bar u/\bar d$ 
     {\it must}  follow the flavor content of the proton, i.e. $\bar u/\bar d \sim 2$ and $\bar u >\bar d$.
     Our data agree with this expectation. 
     
     The  present study has an exploratory nature, and its main goal was to  develop techniques 
     for lattice extraction of PDFs
      based on the ideas of Ref. \cite{Radyushkin:2017cyf}. 
      Our results indicate that the basic method we put forward has a strong potential for obtaining
      reliable PDFs from lattice QCD. In future work we will refine our methods for incorporating evolution and controlling residual polynomial $z_3^2$ effects in the extraction of the Ioffe time distributions.  
      
      To achieve this, 
      it is evident that smaller lattice spacings are required  as  well as a  larger range of nucleon  momenta.  Furthermore, we need to study finite volume effects as well
      as to  incorporate dynamical fermions with pion masses closer to the physical point. We plan to  address all these issues in our future work.

\acknowledgements


One of us (A.R.) thanks V. Braun and X. Ji for discussions and comments. 
We are very grateful to Nobuo Sato  who performed the numerical evolution of the data as described in Sec. V 
 and provided Fig. \ref{evolved}, 
and we are also indebted to him for the help in comparison of our results with global fits. 
This work is supported by Jefferson Science Associates,
 LLC under  U.S. DOE Contract \#DE-AC05-06OR23177. 
 K.O. was supported in part by U.S.  DOE grant
\mbox{ \#DE-FG02-04ER41302}, and A.R. was   supported in part 
 by U.S. DOE Grant   \mbox{\#DE-FG02-97ER41028. }
 S.Z. acknowledges support
by the National Science Foundation (USA) under grant
PHY-1516509. 
 This work was performed in part using computing facilities at the
  College of William and Mary which were provided by contributions from the 
  National Science Foundation (MRI grant PHY-1626177), the Commonwealth of 
  Virginia Equipment Trust Fund and the Office of Naval Research. In addition,  
  this work used resources at NERSC, a DOE Office of Science User Facility supported by the Office of Science of the U.S. Department of Energy under Contract \#DE-AC02-05CH11231.

\end{document}